\documentclass[sn-nature]{sn-jnl}%
\usepackage{graphicx}%
\usepackage{multirow}%
\usepackage{amsmath,amssymb,amsfonts}%
\usepackage{amsthm}%
\usepackage{mathrsfs}%
\usepackage[title]{appendix}%
\usepackage{xcolor}%
\usepackage{textcomp}%
\usepackage{manyfoot}%
\usepackage{booktabs}%
\usepackage{algorithm}%
\usepackage{algorithmicx}%
\usepackage{algpseudocode}%
\usepackage{listings}%
\usepackage{adjustbox}%

\theoremstyle{thmstyleone}%
\theoremstyle{thmstyletwo}%

\theoremstyle{thmstylethree}%
\newcommand{\0}{\phantom{0}}

\begin{document}

\title[Article Title]{Machine Learning on Multiple Topological Materials Datasets}

\author[1,2]{\fnm{Yuqing} \sur{He}}
\author[1]{\fnm{Pierre-Paul} \sur{De Breuck}}
\author*[2]{\fnm{Hongming} \sur{Weng}}\email{hmweng@iphy.ac.cn}
\author[1]{\fnm{Matteo} \sur{Giantomassi}}
\author*[1,3,4]{\fnm{Gian-Marco} \sur{Rignanese}}\email{gian-marco.rignanese@uclouvain.be}

\affil[1]{\orgdiv{Institute of Condensed Matter and Nanosciences}, \orgname{UCLouvain}, \orgaddress{\street{Chemin des Étoiles 8}, \postcode{1348} \city{Louvain-la-Neuve}, \country{Belgium}}}
	
\affil[2]{\orgdiv{Beijing National Laboratory for Condensed Matter Physics and Institute of Physics}, \orgname{Chinese Academy of Sciences}, \orgaddress{\city{Beijing}, \country{China}}}

\affil[3]{\orgname{WEL Research Institute}, \orgaddress{\street{Avenue Pasteur 6}, \postcode{1300} \city{Wavre}, \country{Belgium}}}

\affil[4]{\orgdiv{School of Materials Science and Engineering}, \orgname{Northwestern Polytechnical University}, \orgaddress{\street{No. 127 Youyi West Road}, \city{Xi’an} \postcode{710072} Shaanxi, \country{China}}}

\abstract{%
A dataset of 35,608 materials with their topological properties is constructed by combining the density functional theory (DFT) results of Materiae and the Topological Materials Database.
Thanks to this, machine-learning approaches are developed to categorize materials into five distinct topological types, with the XGBoost model achieving an impressive 85.2\% classification accuracy.
By conducting generalization tests on different sub-datasets, differences are identified between the original datasets in terms of topological types, chemical elements, unknown magnetic compounds, and feature space coverage.
Their impact on model performance is analyzed.
Turning to the simpler binary classification between trivial insulators and nontrivial topological materials, three different approaches are also tested.
Key characteristics influencing material topology are identified, with the maximum packing efficiency and the fraction of \textit{p} valence electrons being highlighted as critical features.
}



\maketitle

\section{Introduction}
 
In the past decade, topological electronic materials have drawn considerable attention due to their importance for both fundamental science and next-generation technological applications~\cite{kitaev2009periodic, hasan2010colloquium, qi2011topological}.
These materials exhibit unique topological configurations in their electronic band structures, resulting in peculiar electronic properties~\cite{hasan2010colloquium, qi2011topological}.
A central and long-standing question in this field is how to determine whether a given material is topologically trivial or not. Thanks to advancements in symmetry indicator theory~\cite{po2017symmetry} and topological quantum chemistry theory~\cite{Bradlyn_2017} and to the use of first-principles calculations~\cite{Zhang_2019, Vergniory_2019, Tang_2019}, it is now possible to categorize topologically non-trivial materials (NTMs) according to their symmetry indicators or elementary band representations (EBRs) and compatibility relations.
One first distinguishes the topological semimetals (TSMs).
These materials present electronic bands intersecting at discrete points or along lines in momentum space, that do not satisfy the compatibility relations, resulting in gapless states near the Fermi level.
These topological nodes cannot be removed by symmetry-preserving perturbations.
Depending on the position of these nodes identified according to the symmetry representations of the corresponding bands, TSMs can be further grouped into high-symmetry-point semimetals (HSPSMs) and high-symmetry-line semimetals (HSLSMs)~\cite{Zhang_2019}.
Depending on the symmetry indicators, one then identifies the topological insulators (TIs) and the topological crystalline insulators (TCIs), which have a set of valence bands that satisfy the compatibility relations but cannot be decomposed into linear combination of EBRs~\cite{Bradlyn_2017}.
Using high-throughput calculations integrating density functional theory (DFT)~\cite{Hohenberg, Kohn} and topological quantum chemistry theory or symmetry indicator theory, tens of thousands of topological materials were detected by analyzing the symmetry of the wavefunctions of crystalline compounds from the Inorganic Crystal Structure Database~\cite{Mariette} (ICSD), leading to the compilation of several databases~\cite{Vergniory_2019, Zhang_2019, Tang_2019}.
Soon after, these two theories based on symmetry representations of wavefunctions have been extended to magnetic ordered materials~\cite{Watanabe2018mag_SI, Elcoro2021mag_TQC, Peng2022mag_SI}.

However, from an experimental standpoint, the main characteristics of materials that may affect their topological properties, especially those which provide helpful chemical insights, are still unclear, hindering the design of new topological materials.
Machine learning (ML)~\cite{samuel1959some} techniques offer a novel approach to address these shortcomings.
By exploring existing data, they can potentially identify the features that are critical for obtaining topological properties without the need to resort to heavy computations.
In this framework, several studies have already been conducted.
For example, neural networks and k-means clustering have been used to learn from the Hamiltonian~\cite{zhang2017quantum, zhang2018machine, zhang2020interpreting, scheurer2020unsupervised}.
Compressed-sensing~\cite{2006Compressed}, gradient boosted trees (GBTs)~\cite{friedman2001greedy}, and other methods have also been employed to learn from ab-initio data~\cite{acosta2018analysis, claussen2020detection, cao2020artificial, liu2021screening, schleder2021machine, andrejevic2022machine, ma2023topogivity}.
Claussen et al.~\cite{claussen2020detection} trained a GBT model with 35,009 symmetry-based entries from the Topological Materials Database (TMD)~\cite{Vergniory_2019, vergniory2022all}.
By testing the effect of different features, their model reached an accuracy of 87.0\% for classifying materials into 5 subclasses: Trivial Insulator (TrIs), two types of TSMs (Enforced Semimetals (ESs) or Enforced Semimetals with Fermi Degeneracy (ESFDs)), and two types of TIs (Split EBRs (SEBRs) or Not a Linear Combination of EBRs (NLC)).
Andrejevic et al.~\cite{andrejevic2022machine} utilized 16,458 inorganic materials from TMD to develop a neural network classifier capable of distinguishing NTMs and TrIs based on X-ray absorption near-edge structure (XANES) spectra. It achieved a $F_1$ score of 89\% for predicting NTMs from their XANES signatures.
 
In 2023, Ma et al.~\cite{ma2023topogivity} introduced a parameter named \textit{topogivity} for each chemical element measuring its tendency to form topological materials.
They employed support vector machine (SVM)~\cite{cortes1995support} to learn the topogivities of elements based on a dataset of 9,026 materials from Tang et al.~\cite{Tang_2019}.
This approach achieved an average accuracy of 82.7\% in an 11$\times$10-fold NCV procedure.
Claussen et al.~\cite{claussen2020detection} found that the topology does not depend much on the particular positions of atoms in the crystal lattice.
This is in contrast with the previous two findings indicating that the local environment of atoms in compounds is a decisive factor since both XANES and topogivity are sensitive to the elements and their local chemical environment.
In addition, these previous studies used mainly generic ML algorithms, overlooking newly designed ones specifically tailored for materials science and which have demonstrated excellent performance~\cite{Chen_2019, Dunn_2020, De_Breuck_2021_1, xie2018crystal}.
Finally, we would like to point out that the outcomes of these studies are difficult to compare due to the diverse range of crystal systems considered (including specific classes of systems, 2D materials, or bulk 3D materials) and the variations in utilized databases, material classes, and types of features incorporated in the ML model construction.
 	
In this paper, we conduct data curation to compile a comprehensive dataset consisting of 35,608 entries from Materiae~\cite{Zhang_2019} and TMD.
This new dataset is then used to train models for classifying a material into five types as TrI, or NTM, specifically HSPSM, HSLSM, TI, and TCI.
Using nested cross-validation (NCV)~\cite{Stone1974} on the data from Materiae, we first benchmark five different approaches: namely Random Forest (RF)~\cite{ho1995random}, XGBoost~\cite{chen2016xgboost} (an implementation of GBTs), Automatminer (AMM)~\cite{Dunn_2020}, the Material Optimal Descriptor Network (MODNet)~\cite{De_Breuck_2021_1,De_Breuck_2021_2}, and the Materials Graph Network (MEGNet)~\cite{Chen_2019}.
We find that XGBoost performs the best.
We then test this model on the complete dataset, achieving a mean NCV accuracy of 82.9\% for the classification into the five types.
We discuss the NCV results and the train-test procedures, highlighting differences between the datasets that affect the scores.
Finally, we compare XGBoost with topogivity~\cite{ma2023topogivity} and t-distributed Stochastic Neighbor Embedding (t-SNE)~\cite{van2008visualizing} for the binary classification between TrIs and NTMs.
XGBoost performs the best with 92.4\% accuracy.
Additionally, we investigate the key factors influencing the topology of materials.
Our analysis reveals that the maximum packing efficiency (MPE) and the fraction of $p$ valence electrons (FPV) are the factors that contribute the most to the distinction between TrIs and NTMs. It is noted that MPE is a structure-based feature that represents the maximum packing efficiency of atoms within a crystal lattice and FPV is a composition-based feature that indicates the fraction of $p$ electrons versus all valence electrons. We think this finding is reasonable and these features could be used as a heuristic for exploring new topological materials.\label{intro}
\section{Results}

\subsection{Data Curation}

Two datasets are constructed in this work, named $M$ and $T$.
The dataset $M$ is extracted from Materiae following a thorough data curation procedure as described below, resulting in 25,683 compounds.
Similarly, the dataset $T$ is constructed from the Topological Materials Database, resulting in 24,156 compounds after cleaning.

\begin{figure}[h]
	\centering
	\includegraphics[width=0.9\textwidth]{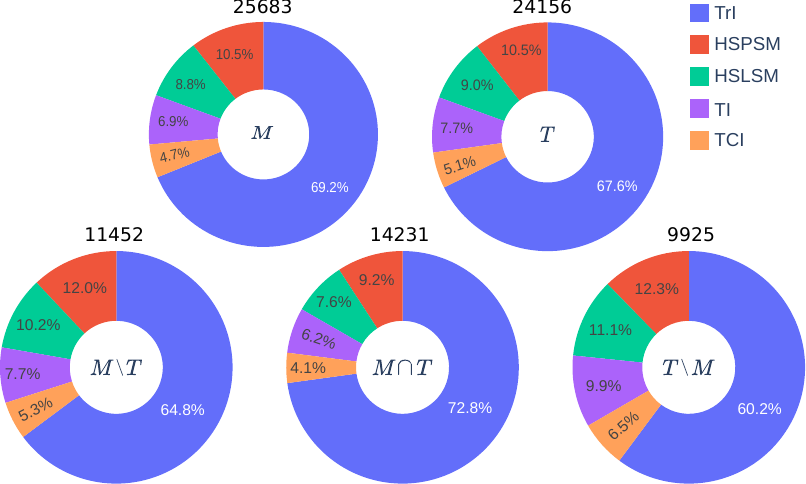}
	\caption{Composition in terms of the different topological types of the datasets $M$ (constructed from Materiae) and $T$ (originating from the Topological Materials Database) as well as of their differences ($M\backslash T$ and $T\backslash M$) and intersection ($M\cap T$).}
	\label{fig:pie5}
\end{figure}

It should be noted that the names of the topological types in these two databases are different.
Therefore, we here establish a correspondence between them during the curation process.
Figure~\ref{fig:pie5} depicts the type distribution for the two datasets, their intersection ($M\cap T$), and differences ($M\backslash T$ and $T\backslash M$).
Roughly the same distribution is found, with a majority of TrIs and around 30\% NTMs.

The data curation proceeds as follows.
For the dataset $M$, we initially query Materiae, which includes 26,120 materials that are neither magnetic materials (i.e., for which the magnetic moment would be higher than 0.1 
Bohr Magneton 
per unit cell according to the Materials Project~\cite{Jain} (MP) record) nor conventional metals (i.e., systems with an odd number of electrons per unit cell).
By keeping only the results that were computed including spin-orbit coupling, an initial dataset of 25,895 materials (named \textit{MAT}) is obtained including their topological properties.
Subsequently, the dataset $M$ is constructed by removing those materials with labels conflicting with the dataset $T$ (see last paragraph). All the records in \textit{MAT} are indexed by their unique MP-ID.

For the dataset $T$, we start from the data available in the Topological Materials Database~\cite{vergniory2022all}, which includes 73,234 compounds indexed by their ICSD-ID and grouped into 38,298 unique materials by common chemical formula, space group, and topological properties as determined from their calculated electronic structure.
As some of the pre-assigned MP-IDs were found to be wrong, we decided to control them systematically with the \textit{structure\_matcher} of \textsc{pymatgen}~\cite{ONG2013314} (using its default tolerance settings) and make our own MP-ID assignment.
Given a set of compounds grouped as one unique material, we distinguish three cases to assign the MP-ID and the corresponding structure.
First, when none of the compounds has an assigned MP-ID, one structure of the set is randomly selected and the corresponding MP-ID is indicated as not available.
Second, when the compounds are associated with at most one MP-ID, one structure of the set is again randomly selected.
We then check whether it matches the MP structure corresponding to the indicated MP-ID.
If it does, the MP-ID is assigned to the structure.
If not, the MP-ID is indicated as not available.
Finally, when more than one MP-ID appears in the set, these MP-IDs are first ranked according to their energy above hull.
Then, the different structures in the set are compared with the MP structures corresponding to these MP-IDs starting from the lowest in energy.
In case of match, the corresponding MP-ID (i.e., the one with the lowest energy) is assigned to the structure.
In case of absence of match with any of the ranked MP-IDs, the MP-ID is indicated as not available.
At the end of the process, only one of the structures associated with the same MP-IDs is kept.
The compounds are then sorted adopting the same classification as in Materiae.
First, they undergoes the same curation as the one described for the dataset M: excluding magnetic materials and conventional metals.
The materials containing rare-earth elements (Pr, Nd, Pm, Sm, Tb, Dy, Ho, Er, Tm, Yb, Lu, Sc) are also removed.
This is done because, for these elements, the results of Materiae and the Topological Materials Database were obtained from calculations performed using pseudopotentials with a different number of valence electrons (typically odd in one case and even in the other).
Furthermore, we label the resulting data according to Materiae’s definition.
For TSMs, the mapping is rather simple: ESFDs correspond to HSPSMs and ESs to HSLSMs. 
In contrasts, for TIs, the mapping is more comple.
We label SEBRs and NLCs as TIs or TCIs as follows.
The materials in the spacegroups 174, 187, 189, 188, or 190 are all labeled as TCIs.
The others are labeled according to the parity of the last topological indices, odd ones as TIs while even ones as TCIs.
The next curation step consists in removing the materials with duplicate MP-IDs, as well the 673 compounds with the same MP-ID but conflicting topological types.
At the end, we were left with a total of 24,368 items with an assigned MP-IP (sometimes indicated as not available) and sorted according to the same classification as Materiae.
Thanks to the curation performed, the compounds in the two datasets can easily be related based on their assigned MP-IDs.
On this basis, we further removed 212 materials present in both datasets but with differing types, leaving 24,156 compounds in $T$.

At the end of the construction of the datasets $T$ and $M$, our global dataset $M\cup T$ contains a total of 35,608 materials while the intersection $M\cap T$ consists of 14,231 compounds, as shown in Fig.~\ref{fig:pie5}.
    
\subsection{Model}

In order to select a model for further training and analysis, we first perform a benchmark on the \textit{MAT} dataset.
Five different models are used: two generic ML algorithms (RF and XGBoost), and three well-developed algorithms in the field of material science (AMM, MODNet, and MEGNet).
Moreover, for each method, two procedures are considered for the multiclass classification: either a direct multiclass classification (which gives the 5 possible labels as output) or a hierarchical binary classification (multiple models are trained following a tree such that each leaf represents a class).
Figure~\ref{fig:trees} schematically represents these two procedures, with their respective accuracies.
The highest accuracy (85.2\%) is obtained with XGBoost using the direct multiclass classification.
It is therefore used in the remainder of the work on all the data.
More details about the benchmark are provided in Sec.~\ref{a_cl}.

\subsection{Generalization tests}

In principle, the NCV score should provide a comprehensive assessment of how the training model performs on new data.
However, the model trained on the dataset $M$, which shows excellent performance (with a NCV accuracy of 85.7\%), is found not to generalize well on the dataset $T\backslash M$ leading to an accuracy of only 71.8\% (i.e., a decrease of 13.9\%).
Therefore, in order to further investigate the model performance, we perform a series of generalization tests by training on the different datasets at our disposal: $M$, $T$, their differences $M\backslash T$ and $T\backslash M$, their intersection $M\cap T$, their union $M\cup T$ as well as the union of their differences $(M\backslash T)\cup (T\backslash M)$.

\begin{figure}
	\centering
	\includegraphics[width=0.8\textwidth]{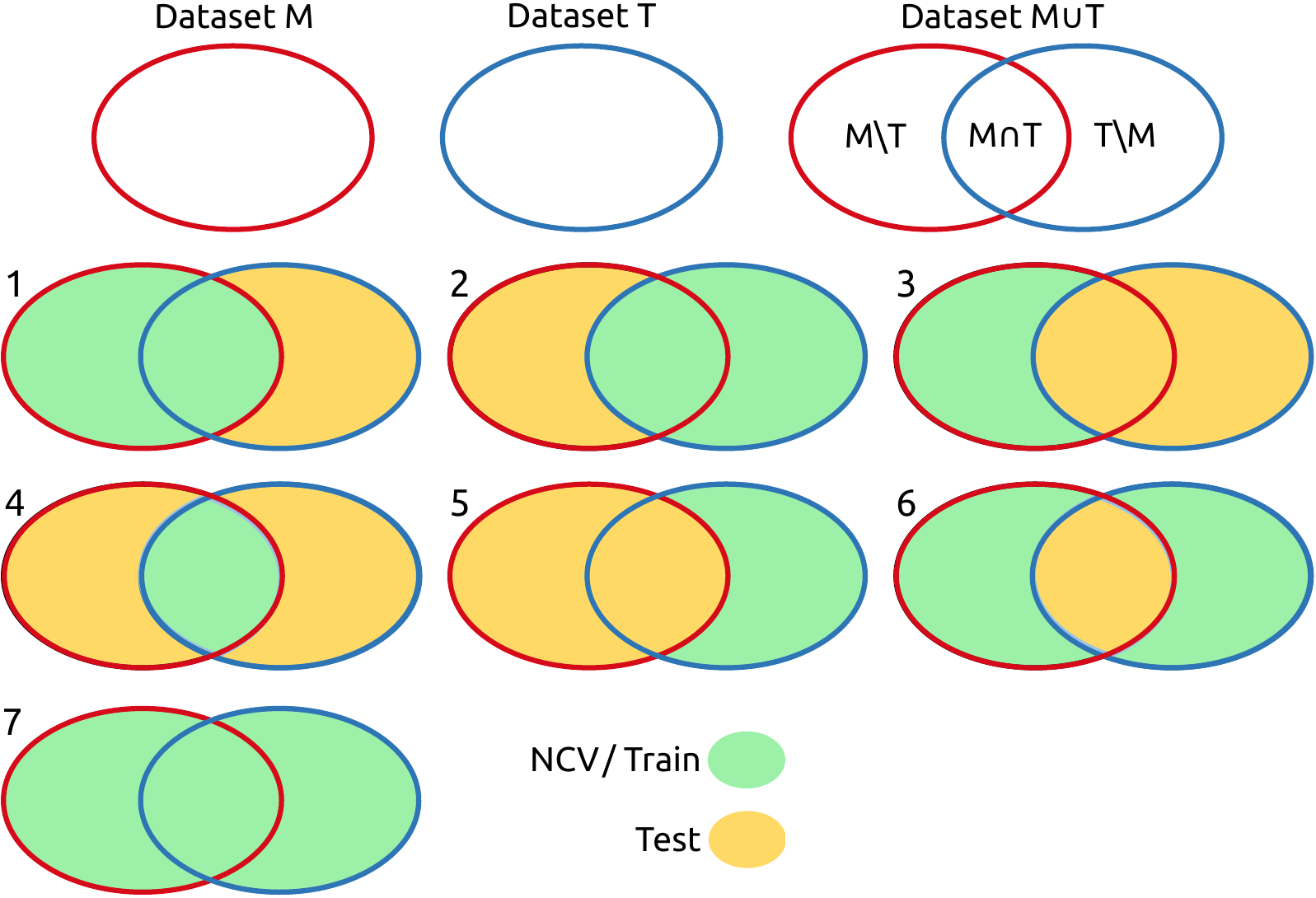}
	\caption{Diagram of the seven generalization tests.
    The datasets $M$ and $T$ are circled in red and blue, respectively.
    The union dataset $M\cup T$ can be split into three part: $M\backslash T$, $M\cap T$, and $T\backslash M$.
    In each of test, the dataset used for the training and the NCV of the ML model is filled in green, while the dataset used for testing it is filled in yellow.}
	\label{fig:Gener_Test}
\end{figure} 
 
The seven tests are schematically represented in Fig.~\ref{fig:Gener_Test}, where the datasets $M$ and $T$ are circled in red and blue, respectively.
In each test, a ML model is first trained on the training set, depicted in green.
A 5-fold NCV test is performed on the same data, followed by a generalization test on the test set depicted in yellow.
The classification accuracy results obtained for each test are reported in Table~\ref{tab:train_test_results1}, indicating the score obtained for the NCV, as well as on $M\backslash T$, $M\cap T$, $T\backslash M$, and $M\cup T$.
Complementary metrics (i.e., $F_1$ score, precision, and recall) are reported in Table~\ref{tab:train_test_results2} in the Appendix.

\begin{table}[]
	\centering
	\caption{Accuracy (in \%) of the different nested cross-validation (NCV) and generalization tests depending on the training dataset.
    The results on (part of) the training dataset evaluated by NCV are indicated by a star.}
	\label{tab:train_test_results1}
    \setlength{\tabcolsep}{6pt}
    \begin{tabular}{lccccccc}
    \hline
    \hline
    Test\textbackslash{}Train & $M$\0 & $T$\0 & $M\backslash T$\0 & $M\cap T$\0 & $T\backslash M$\0 & $(M\backslash T)\cup (T\backslash M)$\0 & $M\cup T$\0  \\  \hline 
    NCV             & 85.7\0 & 80.7\0 & 84.1\0 & 85.1\0 & 72.1\0 & 79.9\0 & 82.9\0 \\
    $M\backslash T$ & 84.8*  & 80.3\0 & 84.1*  & 78.9\0 & 77.1\0 & 85.6*  & 85.8* \\ 
    $M\cap T$       & 86.1*  & 86.0*  & 82.0\0 & 85.1*  & 81.6\0 & 83.6\0 & 86.6* \\
    $T\backslash M$ & 71.8\0 & 73.2*  & 69.8\0 & 70.2\0 & 72.1*  & 73.3*  & 74.3* \\ 
    $M\cup T$       & 81.8\0 & 80.6\0 & 79.3\0 & 79.0\0 & 77.5\0 & 81.4\0 & 82.9\0 \\
    \hline
    \hline
    \end{tabular}
\end{table}

As discussed below, the previously mentioned generalization issue is still present.
For the generalization tests performed with $M$, $T$, $M\backslash T$, and $M\cap T$ as the training set (in green), the NCV accuracy is significantly higher than the test accuracy (i.e., for the corresponding datasets in yellow).
When training on the dataset $M$, the NCV accuracy on the sub-dataset $M\backslash T$ (84.8\%) and $M\cap T$ (86.1\%) is also much larger than the test accuracy (71.8\% for $T\backslash M$).
When training on the dataset $T$, the results are more nuanced with the NCV on the sub-dataset $M\cap T$ (86.0\%) being higher than the test accuracy (80.3\% for $M\backslash T$). But that on the sub-dataset $T\backslash M$ (73.2\%) is not.

It is worth noting that, when training on $T\backslash M$ and $(M\backslash T)\cup (T\backslash M)$, the NCV accuracy (72.1\% and 79.9\%) is smaller than all test accuracy values (81.6\% and 83.6\% for $M\cap T$, respectively; as well as 77.1\% for $M\backslash T$ in the former case).
Finally, the accuracy on $T\backslash M$ is the lowest one whatever the training set.

All the other metrics ($F_1$ score, precision, and recall) reported in Table~\ref{tab:train_test_results2} show the same trend.
All these observations indicate that predicting the topological type on the materials of the dataset $T\backslash M$ seems to be more difficult than on those of the dataset $M$ (or its sub-datasets $M\backslash T$ and $M\cap T$).
We propose four possible explanations for this bias (which are most probably combined).

The first reason is related to the distribution of the topological types in the datasets.
As can be seen in Fig.~\ref{fig:pie5}, the proportion of TrIs is the lowest in $T\backslash M$, and the binary classification between TrIs and NTMs is much more accurate than the subsequent refined classifications of NTMs (see Fig.~\ref{fig:trees}, Node 1 with respect to all the other nodes).
Therefore, the proportion of TrIs affects the global accuracy.
	
The second rationalization is based on the distribution of the chemical elements in the datasets.
Indeed, the accuracy of the model can be very low on compounds containing certain elements (e.g., as low as 37\% on average for Gd), as illustrated in the Appendix (Fig.~\ref{fig:elems_acc}).
In particular, the following elements with a low average accuracy are more present in $T\backslash M$ than in any other dataset: Ne, Mn, Fe, Eu, Gd, Po, Rn, Ra, Am.
To test how this affects the global accuracy in each dataset, we recalculate the performance of the model when these elements are excluded.
The corresponding accuracy, $F_1$ score, precision, and recall as well as the proportion of these materials are reported in the Appendix (Table~\ref{tab:score_elel_proportion_in1}). 
In general, the performance is smaller when including the elements above.
This decrease is more important for the dataset $T\backslash M$ (3\% compared to 0.5\% for the other datasets).
This could be expected as it contains a larger fraction of the elements above.

Following upon this observation, we search for possibly problematic elements in the dataset $M\cup T$.
Their detection is based on more quantitative criteria.
First, the number of materials containing such problematic element should be larger than 30, for statistical reasons.
Second, the accuracy for the compounds containing this element should be lower than 75\%.
Finally, the recall for those materials should be lower than the  one for those without that element.
Applying these criteria, the following elements are identified: Cr, Mn, Fe, Cu, Tc, Eu, Os, Np.
Table~\ref{tab:score_elel_acc} contains the accuracies, $F_1$ score, precision and recall based on the presence of the previous elements.
	
A third potential cause of the bias for the dataset $T\backslash M$ is that about half of its compounds have an unknown magnetic type, since they could not be assigned an MP-ID.
Table~\ref{tab:score_elel_mpid1} investigates both the impact of elements and the presence of magnetic information.
As can be seen, excluding the selected elements in the datasets $M\backslash T$, $M\cap T$ or $T\backslash M$ improves the accuracy by 5.3\%.
Excluding compounds with missing magnetic information further improves the score by 1.2\%.
	
To analyze the cumulative effect of the above three explanations, we define the datasets $\widetilde{M\backslash T}$, $\widetilde{M\cap T}$, and $\widetilde{T\backslash M}$.
These are formed by selecting the same number of compounds (3,372) in each original dataset ($M\backslash T$, $M\cap T$, and $T\backslash M$) adopting the same criteria as in Table~\ref{tab:score_elel_mpid1} and in such a way that the distribution among the five different types is exactly the same (i.e., 2,339 TrIs, 315 HSPSMs, 279 HSLSMs, 271 TIs, and 168 TCIs).
As can be seen in the NCV results reported in Table~\ref{tab:ncv_sample}, the accuracy in the three datasets (79.2\%, 79.9\%, and 77.3\%) is much more similar (the largest difference decreased to 2.6\% from the previous 13.9\%).
	
\begin{figure}
	\centering
	\includegraphics[width=0.45\textwidth]{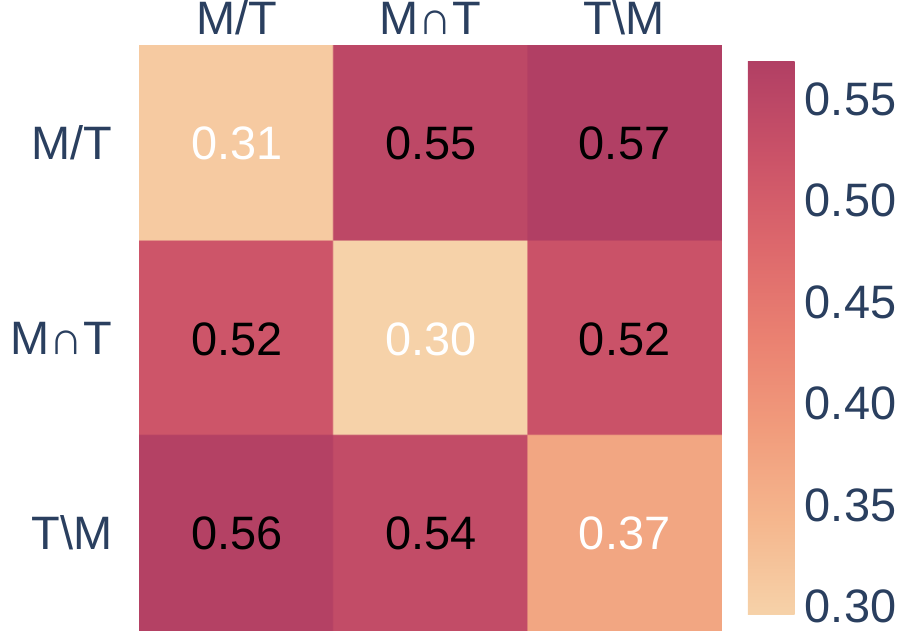}
	\caption{Heterogeneity metric between datasets: $M\backslash T$, $M\cap T$ and $T\backslash M$}
	\label{fig:FD}
\end{figure}

Finally, a fourth possible reason is related to the coverage of the feature space by the datasets.
The ML model performance on a given test set obviously depends on how close its points are from those of the training dataset (interpolative predictions are better than extrapolative ones).
To evaluate this effect, a heterogeneity metric is used, as explained in detail in Methods (see Eq.~\ref{eq:df_AB}).
It quantifies the similarity between the different datasets ($M\backslash T$, $M\cap T$, and $T\backslash M$), with a small heterogeneity leading in principle to a higher performance.
The heterogeneity within each dataset (the diagonal part in Fig.~\ref{fig:FD}) provides a reference value.
Note that the heterogeneity in the dataset $T\backslash M$ is about 20\% larger than in the others.
This may explain the trend in the NCV accuracy for the models trained on the datasets $\widetilde{M\backslash T}$, $\widetilde{M\cap T}$, and $\widetilde{T\backslash M}$: as expected the lower the heterogeneity, the higher the NCV score.
Furthermore, the heterogeneity increases significantly in the off-diagonal elements.
This explains why a model trained on a given dataset tends not to generalize well to the other datasets.
	
\subsection{Binary classification}
	
In order to try to identify the main factors that influence the topology of a material, we turn to the binary classification between TrIs and NTMs on the whole dataset $M\cup T$.
NTMs are considered as positive and TrIs as negative.
Thus, the precision measures the reliability of NTM predictions, and the recall measures the ability to detect all NTMs.
The $F_1$ score, which is the harmonic mean of the precision and the recall, provides a balance between these two quantities (as they typically show an inverse relationship) and offers a better measure than the accuracy for an uneven class distribution.
 
Three approaches are considered here: the XGBoost model as above but for the binary classification; an existing heuristic model based on the \textit{topogivity} of the elements~\cite{ma2023topogivity} relying only on the composition of the compounds; and a generic dimension reduction method t-SNE~\cite{van2008visualizing} applied to the two most important features identified from XGBoost.
All the details are available in Methods.

\begin{figure}
 	\centering
 	\includegraphics[width=0.8\linewidth]{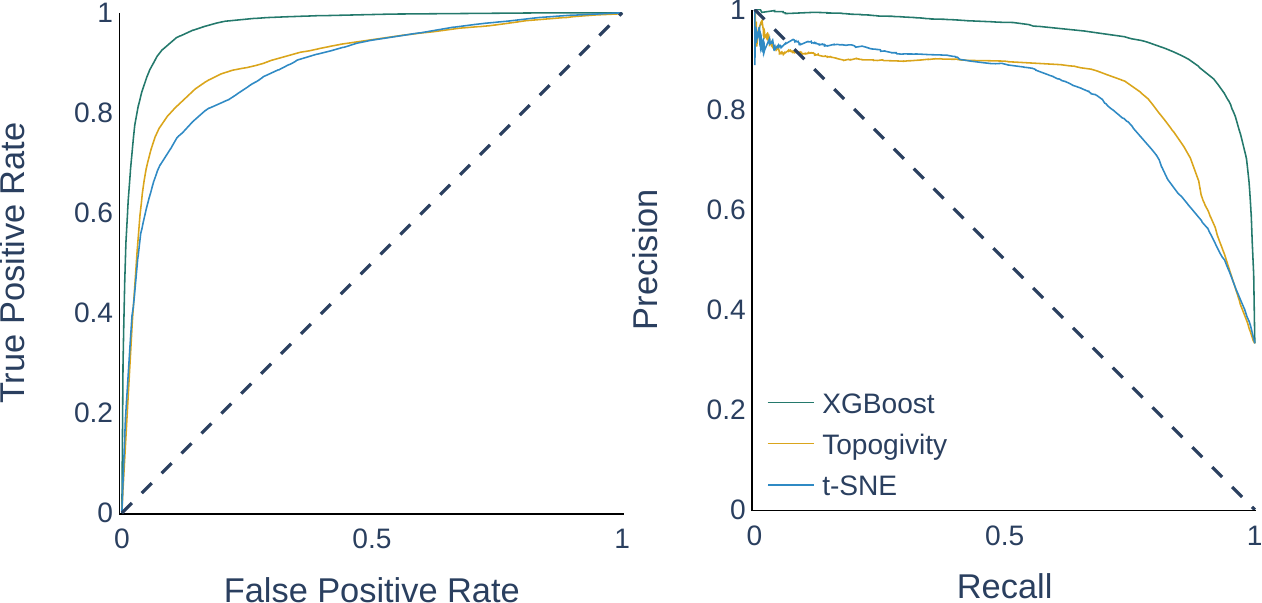}
 	\caption{Receiver operating characteristic (ROC) and precision-recall curves for distinguishing nontrivial topological materials (NTMs) from trivial insulators (TrIs) on the dataset $M\cup T$.}
	\label{fig:all2type3a}
\end{figure}

The results obtained on the dataset $M\cup T$ are provided in Table~\ref{tab:com_mtt} and Fig.~\ref{fig:all2type3a}.
Table~\ref{tab:com_mtt} shows the results of the boolean predictions with the default threshold for each algorithm.

XGBoost shows the best performance with the highest accuracy, $F_1$ score, precision and ROC\_AUC, thanks to its usage of a high-dimensional feature space to represent materials that fully describes the properties of materials.
Figure~\ref{fig:all2type3a} shows the trade-off of the scores as a function of the chosen threshold.
XGBoost always has a better score.
The topogivity and t-SNE approaches present an intersection point where they achieve the same scores.
While their scores are lower than those of XGBoost, the topogivity and t-SNE approaches still provide reasonable results, and their advantage lies in their simplicity, making them easy to interpret.
 
\begin{table}[]
	\centering
	\caption{Comparison of the NCV accuracy, $F_1$ score, precision, and recall (in \%) of the XGBoost, topogivity, and t-SNE approaches.
    The Area Under the Receiver Operating Characteristic Curve (ROC\_AUC) is also reported in \%.}
    \setlength{\tabcolsep}{14pt}
	\begin{tabular}{lccccc}
    \hline
    \hline
	& Accuracy & $F_1$ score & Precision & Recall & ROC\_AUC \\
    \hline
	XGBoost    & 92.4 & 88.5 & 89.5 & 87.5 & 97.5 \\
	Topogivity & 87.2 & 79.5 & 85.6 & 74.1 & 90.6 \\
	t-SNE      & 75.7 & 70.8 & 59.0 & 88.3 & 89.1 \\
    \hline
    \hline
	\end{tabular}
	\label{tab:com_mtt}
\end{table}

\begin{figure}
	\centering
	\includegraphics[width=\textwidth]{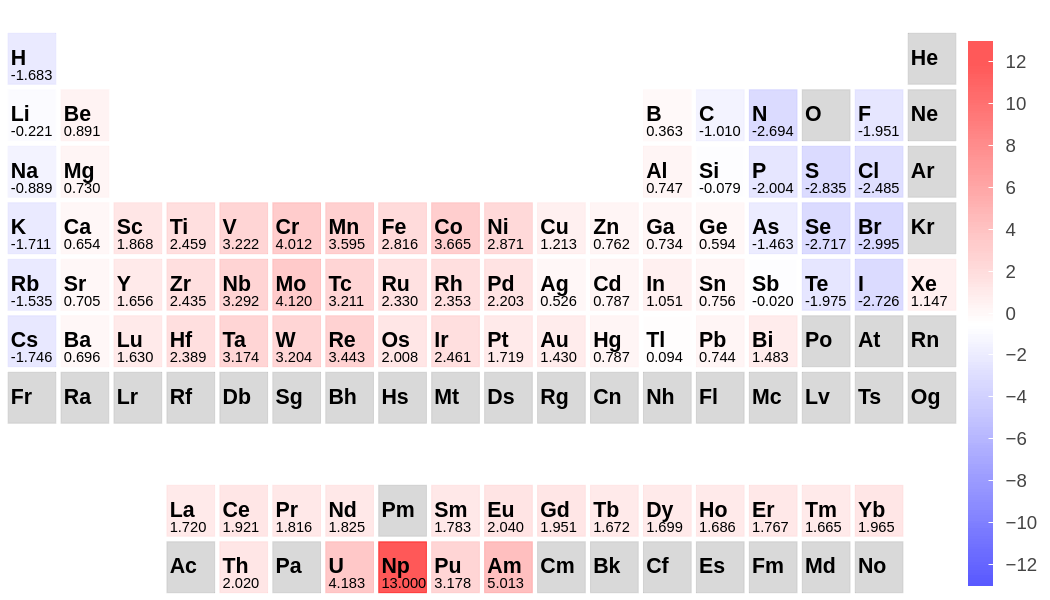}
	\caption{Periodic table of topogivities trained from dataset $M\cup T$.
    Existing topogivities are represented through numerical values with color coding, others are displayed in gray.}
	\label{fig:topogivity}
\end{figure}

The topogivity approach makes predictions based on a simple composition rule (see Eq.~\ref{eq:topog} in Methods) based on a single parameter, the elemental topogivity $\tau_E$ which approximately represents the inclination to form an NTM.
Figure~\ref{fig:topogivity} shows a periodic table with our newly trained topogivities for 83 elements (compared to 54 available previously).

\begin{figure}
	\centering
	\includegraphics[width=0.7\textwidth]{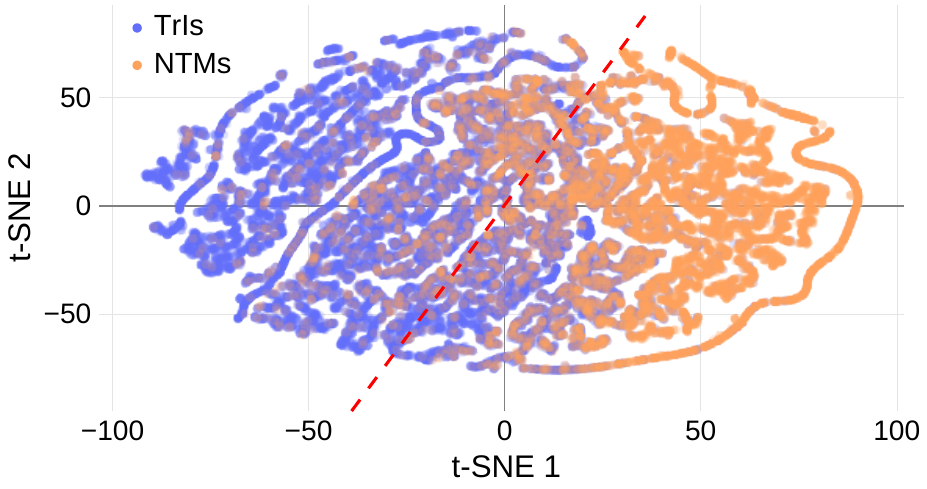}
    \caption{Visualization of the t-SNE results on the dataset $M \cup T$. Non-trivial materials are shown in blue, while trivial insulators are shown in orange. The red dashed line represents the decision boundary obtained using SVM.}

	\label{fig:tsne}
\end{figure}

The t-SNE approach developed here focuses on two features: the maximum packing efficiency in \% (MPE)~\cite{Ward2017} and the fraction of $p$ valence electrons in \% (FPV)~\cite{Ward2016, Deml}.
The points of the whole dataset are represented by two values representing their projections onto the t-SNE variables, as shown in Fig.~\ref{fig:tsne}.
If the points are colored according to their type, a clear separation appears between NTMs and TrIs (in orange and blue, respectively).
Taking the vertical line where t-SNE~1 is equal to zero as the splitting criterion between NTMs and TrIs, it is possible to predict 75.7\% of materials correctly and to detect 88.3\% of the NTMs.
Furthermore, using a soft-margin linear SVM to identify the best frontier (dashed red line), the accuracy reaches 84.7\%.
This is still a bit lower than with the XGBoost and topogivity approaches, but it shows that even without using the target value (hence, in an unsupervised approach), the model can find the underlying relations between features and the topology of materials.
The two selected features are clearly important to determine the topology of materials.

\begin{figure}
	\centering
	\includegraphics[width=\linewidth]{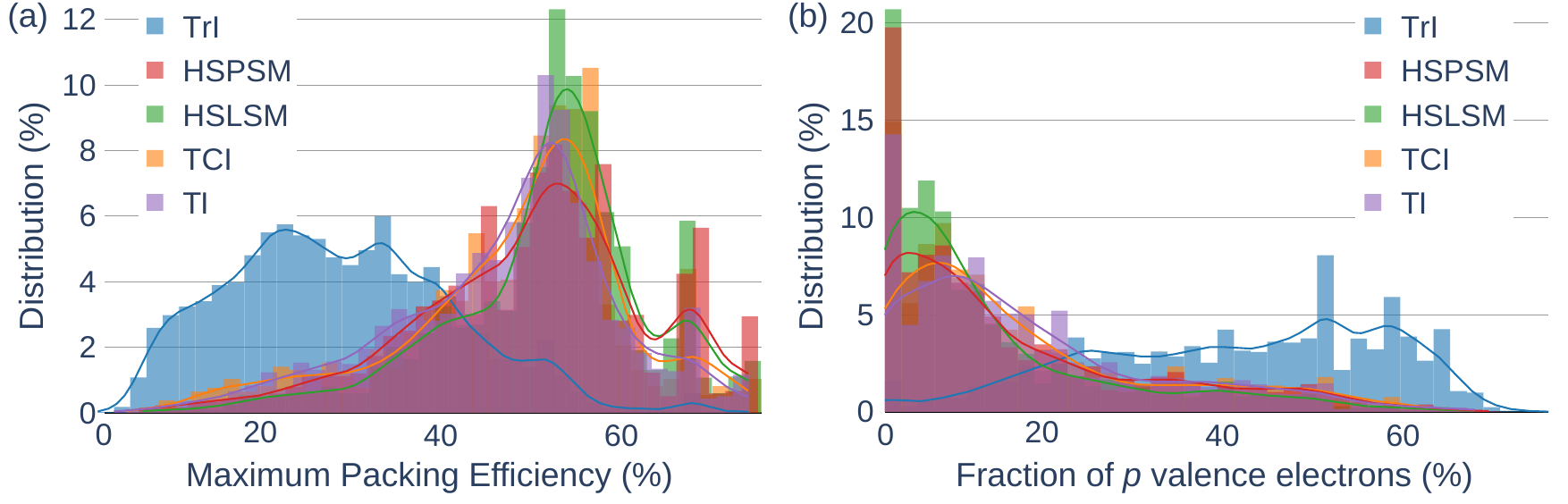}
	\caption{Distinction between trivial insulators (TrI) and nontrivial topological materials (NTM) based on (a) the maximum packing efficiency (\%) and (b) the fraction of $p$ valence electrons in the dataset $M\cup T$.
    The NTM cannot be further discriminated into HSPSM, HSLSM, TCI, and TI.}
	\label{fig:mpe_fpv}
\end{figure}

The distributions of their values in the dataset $M\cup T$ are displayed in Fig.~\ref{fig:mpe_fpv}.
Panel (a) shows that the structures of NTMs are generally more closely packed than TrIs.
This is consistent with our intuition that close-packing structures have stronger interatomic interactions, wider bands, and higher symmetry, thus promoting the appearance of nontrivial topological phases.
Panel (b) demonstrates that NTMs tend to have a lower fraction of $p$ valence electrons.
This can be rationalized as follows.
Compounds with a higher fraction of $p$ valence electrons are mainly composed of elements of the top-right part of the periodic table which are more electronegative.
These tend to form ionic or strongly covalent bonds with a large trivial band gap, hence to generate TrIs.
This observation aligns well with the trends in the element topogivity, as depicted in Fig.~\ref{fig:topogivity}.
Elements located in the top-right part of the periodic table display negative topogivities, indicating their inclination to form TrIs.

\label{results}

\section{Discussion}

In this work, a dataset of 35,608 materials with their topological properties is constructed by combining the DFT results of Materiae and the Topological Materials Database, through a careful cleaning and curation process.
The data from the two databases are found to be generally consistent with only 1\% of the predictions which disagree.
To the best of our knowledge, this is the first integration of materials from distinct data sources, a development that paves the way for more comprehensive and profound machine learning research.
Using this newly created database, two research objectives were pursued.

First, machine-learning approaches were developed for categorizing materials into five distinct topological types (TrI, HSPSM, HSLSM, TI, and TCI).
A thorough benchmark was performed on one of the databases to compare various machine learning approaches, obtained by combining five different models (MEGNet, Automatminer, MODNet, Random Forest, and XGBoost) with two possible procedures, namely one consisting of a series of binary-classification steps to obtain the final sorting, and the other being a direct multiclass categorization.
The direct multiclass procedure relying on XGBoost was identified as the most promising approach, achieving an impressive 85.2\% accuracy.
A series of generalization tests were conducted that allowed for the identification of a series of differences between the two datasets (the distributions of the topological types and the chemical elements in the datasets, the presence of compounds of unknown magnetic type, as well as their coverage of the feature space). 
Their influence on the performance of the model was carefully analyzed.

Secondly, key factors influencing the material topology were identified by focusing on the binary classification between TrIs and NTMs.
The previously developed approach relying on XGBoost performs even better on this simpler task achieving 92.4\% accuracy.
It was compared with two simpler methods, one relying on the use of the \textit{topogivity} of the elements~\cite{ma2023topogivity}, and one being an unsupervised t-SNE.
The latter only focuses on the two features identified as the most important, namely MPE and FPV. 
It demonstrates an accuracy of 84.7\%.
Such performance shows that the two features are very relevant to determining the topology of materials.

Upon analyzing the distribution of these features, we found notable disparities between NTMs and TrIs.
These factors are compatible with our understanding and, together with topogivity, they offer heuristic intuitions for designing topological materials.
This highlights the potential to discover critical features using machine learning approaches.	

Prior to our work, Claussen et al.~\cite{claussen2020detection} had used a similar machine learning approach on TMD, also relying on XGBoost.
They had found that the topology is mostly determined by the chemical composition and the crystal symmetry and that it does not depend much on the particular positions of atoms in the crystal lattice.
In contrast, Andrejevic et al.~\cite{andrejevic2022machine} had suggested that XANES, a spectrum strongly related to the atomic type, the site, and the short-range interactions with surrounding atoms, could be used to identify NTMs and TrIs.
Topogivity had also been introduced~\cite{ma2023topogivity} as an intuitive chemical parameter related only to the elemental composition and had been found to work rather well.
Therefore, it remained elusive whether or not local chemical environment and element-related atomic features are the important characteristics influencing the topology.
In our study, we included new descriptors related to the crystal structure, the composition, and the atomic sites, in addition to those employed by Claussen et al.~\cite{claussen2020detection}, enlarging the space to be explored.
We observed that, rather than the crystal symmetry, the most important characteristic influencing the topology is MPE of atoms in the crystal lattice space.
We rationalize this by the fact that the latter actually determines the hopping parameters of electrons and as a result influences the band width and the possibility of band inversion leading to non-trivial band topology.
We found that the second most important characteristic is FPV.
We think that the latter has an impact on the type of bonding.
Indeed, many compounds showing essentially $p$ orbitals in the valence tend to be large band gap covalent insulators like diamond and silicon, or ionic insulators like NaCl and CaF$_2$.

Our study reveals the critical role of a comprehensive database in the ML research.
The acquisition of a more extensive dataset, encompassing not only symmetry-indicator-based topological materials but also simulation results from Wilson loops, along with experimental data, holds immense importance for driving the further progress of this research.\label{discussion}
\section{Methods}

\subsection{ML Models}

In this study, three main approaches have been considered for the classification of topological materials.
The first one is a ML classifier into the five different types (TrI, HSPSM, HSLSM, TI, and TCI).
For this approach, we tested five different models combined with two possible procedures.
The other two can only separate the materials into two classes, namely TrIs and NTMs (the first approach can obviously also produce this simpler classification).
The second one relies on the use of a previously developed heuristic model based on the \textit{topogivity} of the elements~\cite{ma2023topogivity}.
The last one is an unsupervised ML approach relying on t-SNE.

\subsubsection{ML classifier}
\label{a_cl}

For the ML classifier, we benchmark five different models (MEGNet, Automatminer, Random Forest, MODNet, and XGBoost) with two possible procedures.
These are benchmarked on the dataset \textit{MAT} to determine the best model.

\textbf{ML models}

MEGNet~\cite{Chen_2019} is a graph neural network for machine learning molecules and crystals in materials science.
MEGNet v1.2.9\footnote{\url{https://github.com/materialsvirtuallab/megnet}} is used in this work.
For the multiclass classification (Tree 2, see below), the last layer is changed to softmax and the loss function to "categorical\_crossentropy".
In the MEGNet model, the crystal is represented through a CrystalGraph which is truncated using a cutoff radius of 4\AA \space for defining the neighbors of each atom.
It is trained using 500 epochs.
Given that structures containing isolated atoms 
cannot be handled by MEGNet, they are discarded from the training and test sets.
The scores reported for MEGNet refer to the results obtained on the valid structures (i.e., without isolated atoms).

In all the other ML models, the crystal is represented using the features generated by Matminer~\cite{WARD201860}.
This library transforms any crystal (based on their composition and structure) into a series of numerical descriptors with a physical and chemical meaning.
It relies on various featurizers adapted from scientific publications.

Automatminer (AMM)~\cite{Dunn_2020} allows for the automatic creation of complete machine learning pipelines for materials science.
Here, features are automatically generated with Matminer and then reduced.
Using the Tree-based Pipeline Optimization Tool (TPOT) library~\cite{Olson2016EvoBio}, an AutoML stage is used to prototype and validate various internal ML pipelines.
A customized 'express' preset pipeline is performed on the training set (note that the 'EwaldEnergy' is excluded from the AutoFeaturizer due to technical problems).

Random Forest (RF)~\cite{ho1995random}, which is an ensemble learning method, constructs multiple decision trees.
For the classification task, the final output is the result of majority voting.
Here, we first use Matminer to extract Magpie\_ElementProperty~\cite{Ward2016}, SineCoulombMatrix~\cite{Faber2015}, DensityFeatures and GlobalSymmetryFeatures from the input structures (the missing features are filled with the average of the known data).
The hyperparameters are determined using a 5$\times$3 NCV on the training data, where the stratified 3-fold inner cross-validation uses a grid search for determining the optimal values (\{'max\_features' : ['auto', 'sqrt', 'log2'], 'criterion' : ['gini', 'entropy']\}) relying on the 'balanced\_accuracy' score.

MODNet~\cite{De_Breuck_2021_1,De_Breuck_2021_2} is a framework based on feature selection and a feedforward neural network.
The selection uses a relevance-redundancy score defined from the mutual information between the features and the target and between pairs of features.
The framework is well suited for limited datasets.
Here, we first use a predefined set of featurizers (DeBreuck2020Featurizer with accelerated oxidation state parameters) to generate features from the structures.
Table~\ref{tab:allfeas} provides a complete list of all these Matminer featurizers. 
The missing features are filled with their default value (most are zero).
The feature selection is performed on the training data only.
The model hyperparameters are determined using a genetic algorithm.

\begin{table}[]
	\centering
	\caption{The set of Matminer featurizers to generate the numerical descriptors in MODNet. They are used for algorithm MODNet and XGBoost.}
    
    \begin{tabular}{p{3.2cm}p{3.5cm}p{5.0cm}}
    \hline 
    \hline
    Composition                       & Structure                  & Site                                  \\ \hline
    AtomicOrbitals                    & BondFractions              & AGNIFingerprints                      \\
    AtomicPackingEfficiency           & ChemicalOrdering           & AverageBondAngle                      \\
    BandCenter                        & CoulombMatrix              & AverageBondLength                     \\
    ElectronegativityDiff             & DensityFeatures            & BondOrientationalParameter            \\
    ElementFraction                   & EwaldEnergy                & ChemEnvSiteFingerprint                \\
    ElementProperty  & GlobalSymmetryFeatures     & CoordinationNumber                    \\
     (“magpie” preset)             &  & \\

    IonProperty                       & MaximumPackingEfficiency   & CrystalNNFingerprint                  \\
    Miedema                           & RadialDistributionFunction & GaussianSymmFunc                      \\
    OxidationStates                   & SineCoulombMatrix          & GeneralizedRadialDistributionFunction \\
    Stoichiometry                     & StructuralHeterogeneity    & LocalPropertyDifference               \\
    TMetalFraction                    & XRDPowderPattern           & OPSiteFingerprint                     \\
    ValenceOrbital                    &                            & VoronoiFingerprint                    \\
    YangSolidSolution                 &                            &                      \\  
        \hline
        \hline
	\end{tabular}
	\label{tab:allfeas}
\end{table}

XGBoost~\cite{chen2016xgboost} is an ensemble of boosted trees.
Here, the features generated by the MODNet approach are used as the input of the model.
The hyperparameters are determined using a 5$\times$5 NCV on the training data, where the stratified 5-fold inner cross-validation uses a grid search for determining the optimal values (see Table~\ref{tab:params_xgbt}) relying on the $F_1$ score.

\textbf{Procedures}

\begin{figure}
	\centering
	\includegraphics[width=\textwidth]{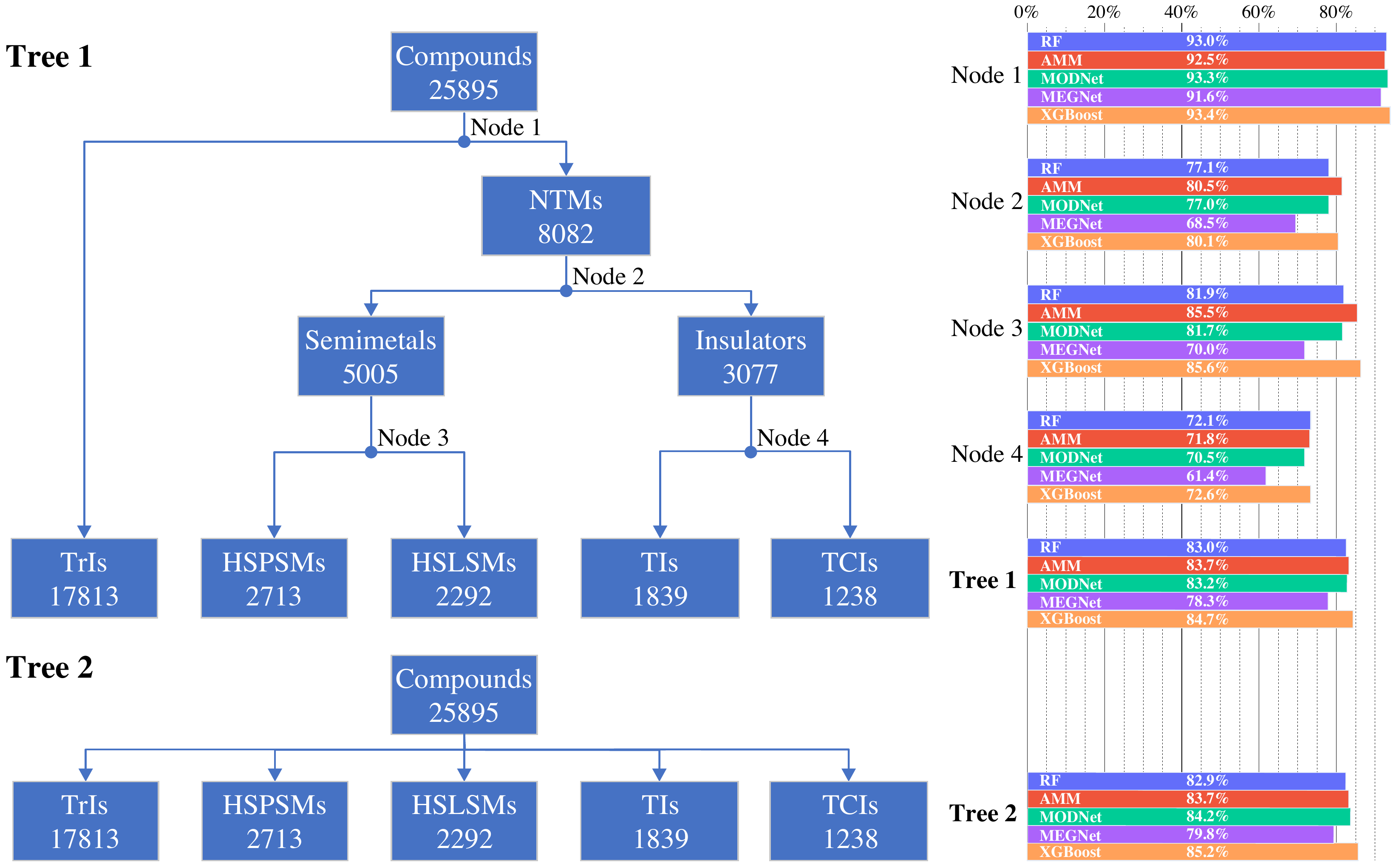}   
	\caption{Illustration of the two different procedures used to categorize the materials according to their topological types.
    The Tree 1 relies on four binary-classification steps, while Tree is base on a direct multiclass approach.
    The accuracy achieved with the five ML models (RF, AMM, MODNet, MEGNet, and XGBoost) are reported both at the final stage of both procedures, as well as at each step in Tree 1.}
	\label{fig:trees}
\end{figure}

Two different classification procedures are used, as shown in Fig.~\ref{fig:trees}.
The first approach (Tree 1) uses four one-vs-all binary-classification steps to obtain the final classification.
The second approach (Tree 2) uses a direct multi-class classification for the five different types, using either majority voting for tree-based methods, or softmax activation for neural network based methods.
The accuracy for both approaches are reported in the figure.

\textbf{Benchmark}

A consistent NCV testing procedure is used to evaluate the performance on the dataset \textit{MAT} for the 10 different combinations of ML models and procedures.
For all the ML models, an identical stratified 5-fold outer loop was used to determine the generalization error (test error).

The internal validation depends on the algorithm, as explained above.

In terms of the final classification, the results of the two procedures are very close with a ranking that depends on the ML model.
The best result is achieved for the direct multiclass procedure relying on XGBoost which achieves an accuracy of 85.2\%.

\begin{table}[]
	\centering
	\caption{Hyperparameter grid searched for the XGBoost model.}
    \setlength{\tabcolsep}{8pt}
	\begin{tabular}{llll}
    \hline
    \hline
	Parameter & Value & Parameter & Value \\
    \hline
	Learning rate $\eta$  & 0.23  & $L^{2}$ regularization $\lambda$ & 1.33 \\
	Maximal tree depth & [9, 10, 11] &  Miniaml child weight & [0.1, 0.3, 0.5] \\ 
	Column subsampling by tree & [0.75, 0.78] & Column subsampling by node & [0.75, 0.78] \\ 
    \hline
    \hline
	\end{tabular}
	\label{tab:params_xgbt}
\end{table}

\subsubsection{Topogivity}
\label{a_tg}
	
The topogivity $\tau_E$ of an element $E$ has been proposed as a measure of its tendency to form topological materials~\cite{ma2023topogivity}.
The $\tau_E$ values for 54 elements were originally obtained by using support vector machine (SVM)~\cite{cortes1995support} model trained on a subset of 9,026 compounds from the database created by Tang \textit{et al.}~\cite{Tang_2019}, with approximately one half classified as trivial and the other half as topological.
The ML model is based on a heuristic chemical rule, which maps each material $M$ to a number $g(M)$ through the function 
\begin{equation}
	g(M) = \sum_{E}f_E(M)\tau_E 
	\label{eq:topog}
\end{equation}
where the summation runs over all elements $E$ in the chemical formula of material $M$, with $f_E(M)$ denoting the fraction of element $E$ within material $M$.
A material $M$ is classified based on $g(M)$, as trivial (TrI) if negative and topological (NTM) if positive.

Here, we also build a new topogivity model trained on a subset from dataset $M\cup T$, which excludes the elements occurring less than 25 times.
We construct a soft-margin linear SVM using the scikit-learn library (specifically, the sklearn.svm.SVC class).
The hyperparameter $C$ of the model is determined through a grid search among 15 values evenly spaced on a log scale ranging from $10^4$ to $10^6$ relying on the $F_1$ score and adopting a 5-fold validation procedure.

The optimal value ($C = 3.73 \times 10^4$) is then used to train the final model on the whole dataset.
This new topogivity model provides the $\tau_E$ value for 83 elements and covers 35,522 materials out of the 35,608 of the dataset $M\cup T$.
These values are reported on the periodic table shown in Fig.~\ref{fig:topogivity}.

For comparison, the previous model~\cite{ma2023topogivity} gives the $\tau_E$ value for 54 elements and covers only 18,637 compounds in $M\cup T$.
A quantitative comparison of the two models can thus only be performed on these 18,367 compounds.
From the different scores reported in Table~\ref{tab:score_topog}, it can be said that their performance is essentially similar.

\begin{figure}
    \centering
    \includegraphics[width=1\textwidth]{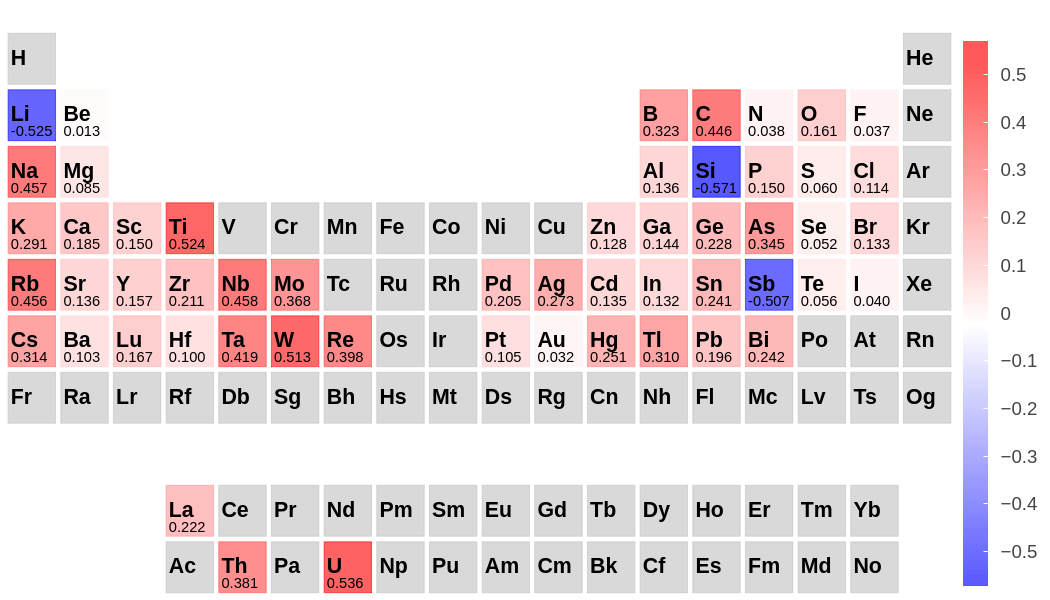}
    \caption{Comparison of the topogivities of Ref.~\cite{ma2023topogivity} with those obtained here for 54 elements.
    The color code refers to $\Delta\tau_E$, a measure of their relative difference, as defined by Eq.~(\protect\ref{eq:topog_diff}).
    When the signs of the two topogivities are different, the value of $\Delta\tau_E$ is negative, and thus the element is highlighted in blue.}
    \label{fig:topog_diff}
\end{figure}

To compare the differences element by element, we use a form of relative difference defined by:
\begin{equation}
    \Delta\tau_E = \frac{2\,\text{sign}(\tau_E^p \tau_E^n) \left |\tau_E^p-\tau_E^n \right | }{2+\left |\tau_E^p \right | +\left | \tau_E^n \right|},
    \label{eq:topog_diff}
\end{equation}
where $\tau_E^p$ and $\tau_E^n$ are the topogivities of the previous and new models, respectively.
The values of $\Delta\tau_E$ are indicated on the periodic table shown in Fig.~\ref{fig:topog_diff}.
They show that the two sets of results are essentially consistent with each other, except for three elements (Li, Si, and Sb), which exhibit different signs (as indicated by the blue color in the figure).

\begin{table}[]
	\centering
	\caption{NCV accuracy, $F_1$ score, precision and recall (in \%) of the two different topogivity models using the 18,637 compounds of the $M\cup T$ dataset which only include the 54 elements for which $\tau_E$ is provided in Ref.~\cite{ma2023topogivity}.}
    \setlength{\tabcolsep}{15pt}
	\begin{tabular}{lccccc}
    \hline
    \hline
	\textbf{}     & Accuracy & $F_1$ score & Precision & Recall    \\
    \hline
	This work                     & 90.2 & 76.8  & 83.5 & 71.2 \\
	Ref.~\cite{ma2023topogivity}  & 89.6 & 77.3  & 77.1 & 77.5\\
    \hline
    \hline
	\end{tabular}
	\label{tab:score_topog}
\end{table}

The final evaluation of the topogivity approach is performed using a 5$\times$5 NCV.
The results are presented in Table~\ref{tab:com_mtt} and Fig.~\ref{fig:all2type3a}.
It leads to an accuracy of 87.2\%, a $F_1$ score of 79.5\%, a precision of 85.6\%, and a recall of 74.1\%.

\subsubsection{t-SNE}
\label{a_tsne}

The last approach originates from the idea to identify the most important features to distinguish topological materials, which we then combine with the unsupervised algorithm t-SNE.
All the Matminer features are first ranked in descending order of gain importance for training the XGBoost classification model between the 5 types for the datasets $M\backslash T$, $M\cap T$ and $T\backslash M$.
The intersection of the top 15 most important features consists of 3 features: the maximum packing efficiency (MPE) in \%, the fraction of $p$ valence electron (FPV) in \%, and the formation enthalpy ($\Delta$H) in eV/atom as obtained from the semi-empirical Miedema model~\cite{miedema_zhang_2016}.

\begin{figure}
	\centering
	\includegraphics[width=0.9\linewidth]{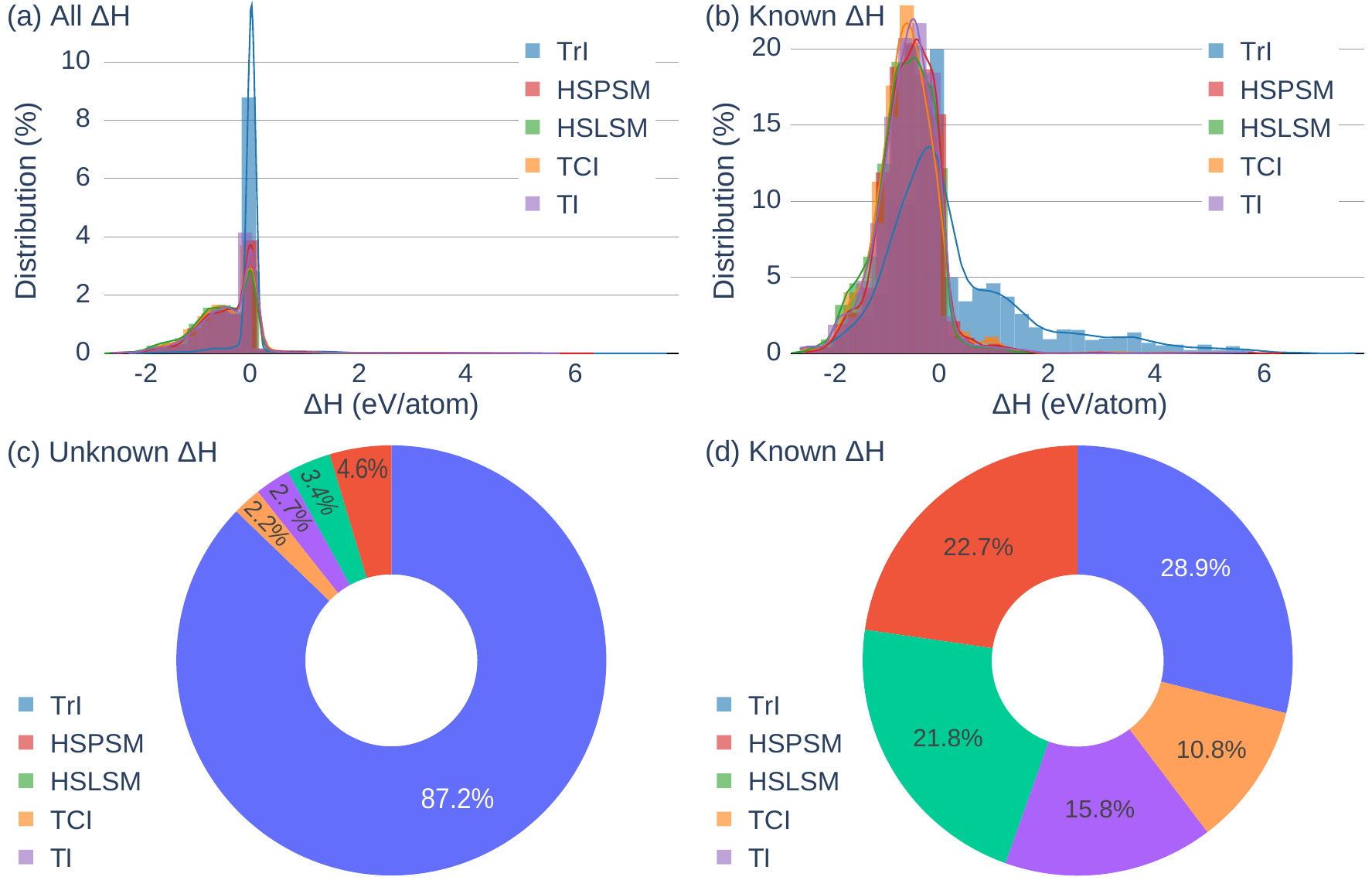}
	\caption{Distribution of $\Delta$H values in the dataset $M\cup T$ according to the 5 topological types.
    Panel (a) shows the distribution of all the values, while panel (b) presents the distribution only for those compounds for which the value is known through the Miedema model~\cite{miedema_zhang_2016}.
    Panels (c) and (d) show the pie-charts with distribution of the compounds among the 5 different topological types for those compounds with unknown and known values, respectively.}
	\label{fig:mdh}
\end{figure}

Based on this finding, the distribution of the compounds in $M\cup T$ according to the values of MPE, FPV, and $\Delta$H is analyzed to understand what helps the models to distinguish between TrIs from NTMs.
The corresponding plots, in which the five different materials types have been separated, are reported in Fig.~\ref{fig:mpe_fpv} for MPE and FPV (already commented in the Results section) and in Fig.~\ref{fig:mdh}(a) for $\Delta$H.

For the latter, it turns out that there is a concentration of TrIs around zero.
The reason for this is, however, not physical but technical.
In fact, the semi-empirical Miedema model~\cite{miedema_zhang_2016} does not apply to all materials.
For those compounds where it does not work, Matminer does not produce a value for the feature and, as commonly done in ML approaches, we replace these missing values with a zero.
As it turns out, as shown in Fig.~\ref{fig:mdh}(c), the share of such missing values is much larger for TrIs (more than 87\%) than NTMs, the ML model takes advantage of this flaw to identify them.

If the compounds without $\Delta$H value are removed, 12,518 entries are left in $M\cup T$, rather evenly sorted into 3,622 TrIs, 2,843 HSPSMs, 2,734 HSLSMs, 1,339 TCIs, and 1,980 TIs (see Fig.~\ref{fig:mdh}(d)).
The corresponding distribution of the $\Delta$H values is reported in Fig.~\ref{fig:mdh}(b).
It still shows a significant difference between TrIs (which mostly present positive values) and NTMs (which are mainly negative).
The underlying reasons are still not completely clear to us.
Given that value of the feature is missing for many materials, it can, however, not be used as a discriminator.
Therefore, it is discarded from the subsequent analysis.

\begin{figure}
	\centering
	\includegraphics[width=0.7\textwidth]{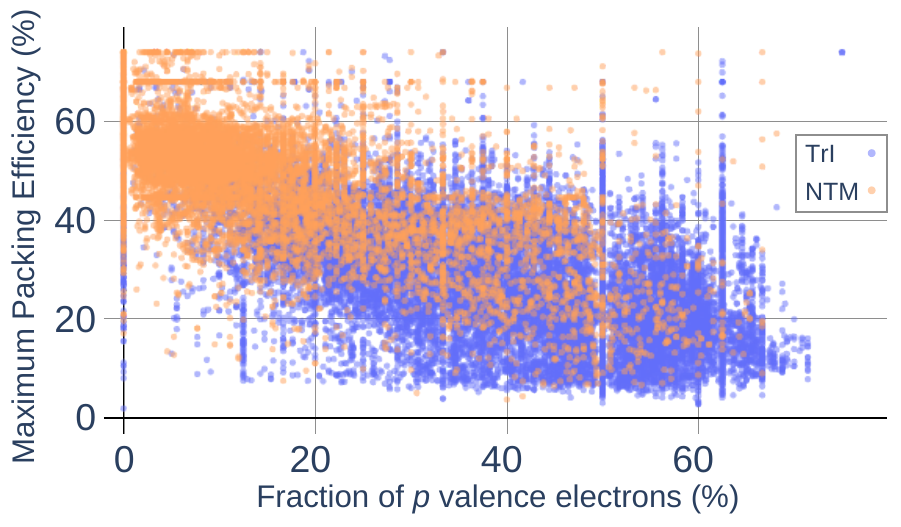}
	\caption{Visualization of the dataset $M\cup T$ according to the two most important features: the maximum packing efficiency and the fraction of $p$ valence electrons.
    This representation leads to a clear separation between TrIs and NTMs even though there is some overlap between the two types.}
	\label{fig:2feas}
\end{figure}

The dataset $M\cup T$ can now be simply visualized in 2D by representing each material by its values for MPE and FPV, as shown in Fig.~\ref{fig:2feas}.
The plot reveals a rather clear distinction between TrIs and NTMs.
And, if we apply t-SNE on the whole dataset, the distinction is even clearer.
Finally, based on the t-SNE variables, a dividing line can be drawn using a soft-margin linear SVM in which $C$ is equal to 1.0.

\subsection{Heterogeneity metric}
\label{a_fd}
	
To quantify the degree of diversity within a dataset or between two datasets, we adopted a heterogeneity metric defined as the k-nearest-neighbor (k-NN) distance in the space of the most important Matminer features.
The most important features are defined based on their induced gain during the training of the XGBoost model.
A total of 47 Matminer features (44 presenting continuous values and 3 discrete ones, respectively) was gathered by taking the union of the 20 most important ones for training on the sub-datasets $M\backslash T$, $M\cap T$, and $T\backslash M$.
These originate from the five featurizers reported in Table~\ref{tab:fd47}.
 
\begin{table}[]
	\centering
	\caption{Matminer featurizers generating the features selected for defining the heterogeneity metric between datasets.
    A brief description and the source of the data is also provided.}
    \setlength{\tabcolsep}{3pt}

	\begin{tabular}{p{3.8cm}p{8cm}}
    \hline
    \hline
	Matminer Featurizer              &  Description \\ 
    \hline
	$\bullet$ Miedema                & Formation enthalpies of intermetallic compounds~\cite{miedema_zhang_2016}.\\
    $\bullet$ ElementProperty        & Weighted elemental statistics~\cite{Ward2016}.\\
    \0 (“magpie” preset)             & \\
	$\bullet$ OxidationStates        & Statistics about the oxidation states for each specie.
    Features are concentration-weighted statistics of the oxidation states.\\
	$\bullet$ AtomicOrbitals         & Estimation of the highest occupied molecular orbital (HOMO) and lowest unoccupied molecular orbital (LUMO) energies based on the composition and the atomic orbital energies~\cite{Svetlana}.\\ 
	$\bullet$ GlobalSymmetryFeatures & Symmetry information.\\
    \hline
    \hline
	\end{tabular}
	\label{tab:fd47}
\end{table}	

All the continuous feature values are rescaled to range between 0 to 1 using the MinMaxScaler from scikit-learn on the dataset $M\cup T$.
Then, the distance $d(x,y)$ between any two points $x$ and $y$ is defined as:
\begin{equation}
	d(x,y) = \sqrt{\sum_{i=1}^{n_c}(x_i-y_i)^{2}+\sum_{j=n_c+1}^{n_c+n_d}(1/2(x_j==y_j))^{2}} \  ,
	\label{eq:df_euc}
\end{equation}
where $n_c$=44 and $n_d$=3 are the numbers of features presenting continuous and discrete values, respectively.
The distance $D(x,A)$ between a point $x$ and a dataset $A$ is defined as the mean of the distances from point $x$ to the 5 NN-points ($q_j$ with $j=1, \ldots, 5$) in dataset $A$:
\begin{equation}
    D(x,A) = \frac{1}{5}\, \sum_{j=1}^{5} d(x,q_j).
\end{equation}
Finally, the heterogeneity metric $H(A,B)$ between two datasets $A$ and $B$ (which can be the same dataset $A$) is defined based on the average distance between all the points of $A$ ($p_i$ with $i=1, \ldots, N_A$) and the dataset $B$:
\begin{equation}
	H(A,B) = \frac{1}{N_A} \sum_{i=1}^{N_A} D(p_i,B).
		\label{eq:df_AB}
\end{equation}
Note that with this definition $H(A,B)$ is an asymmetric measure (it follows from the fact that training and testing sets are not interchangeable).\label{method}
\section{Data availability}

The data can be downloaded from the Materials Cloud Archive~\cite{Talirz2020}: \url{https://doi.org/10.24435/materialscloud:xx-xb}.
It can be viewed interactively (e.g., the plots corresponding to Figures~\ref{fig:mpe_fpv} and \ref{fig:2feas}) through the chemiscope visualization tool~\cite{Fraux2020}.
A Jupyter notebook is also available.
All this information is also available here: \url{https://topoclass.modl-uclouvain.org}.

\backmatter

\bmhead{Acknowledgments}

We thank N. Regnault for providing the data of TMD, which makes the data curation of the Materiae and TMD possible. Y. H. and H. W. acknowledge the funding from the National Key Research and Development Program of China (Grant No. 2022YFA1403800), the National Natural Science Foundation of China (Grants No. 12188101, 11925408), the Chinese Academy of Sciences (Grant No. XDB33000000). Y. H. was supported by China Scholarship Council (Grant No. 201904910878). H. W. is also supported by the New Cornerstone Science Foundation through the XPLORER PRIZE. Computational resources have been provided by the supercomputing facilities of the Université catholique de Louvain (CISM/UCL) and the Consortium des Équipements de Calcul Intensif en Fédération Wallonie Bruxelles (CÉCI) funded by the Fond de la Recherche Scientifique de Belgique (F.R.S.-FNRS) under convention 2.5020.11 and by the Walloon Region.

\bibliography{article}

\begin{thebibliography}{10}
\expandafter\ifx\csname url\endcsname\relax
  \def\url#1{\burl{#1}}\fi
\expandafter\ifx\csname urlprefix\endcsname\relax\def\urlprefix{URL }\fi
\providecommand{\bibinfo}[2]{#2}
\providecommand{\eprint}[2][]{\url{#2}}
\providecommand{\doi}[1]{\url{https://doi.org/#1}}
\bibcommenthead

\bibitem{kitaev2009periodic}
\bibinfo{author}{Kitaev, A.}
\newblock \bibinfo{editor}{{\color{white}a}} (ed.) \emph{\bibinfo{title}{Periodic table for topological insulators and superconductors}}.
\newblock (ed.\bibinfo{editor}{{\color{white}a}}) \emph{\bibinfo{booktitle}{AIP conference proceedings}}, Vol. \bibinfo{volume}{1134}, \bibinfo{pages}{22--30} (\bibinfo{organization}{American Institute of Physics}, \bibinfo{year}{2009}).

\bibitem{hasan2010colloquium}
\bibinfo{author}{Hasan, M.~Z.} \& \bibinfo{author}{Kane, C.~L.}
\newblock \bibinfo{title}{Colloquium: topological insulators}.
\newblock \emph{\bibinfo{journal}{Rev. Mod. Phys.}} \textbf{\bibinfo{volume}{82}}, \bibinfo{pages}{3045} (\bibinfo{year}{2010}).

\bibitem{qi2011topological}
\bibinfo{author}{Qi, X.-L.} \& \bibinfo{author}{Zhang, S.-C.}
\newblock \bibinfo{title}{Topological insulators and superconductors}.
\newblock \emph{\bibinfo{journal}{Rev. Mod. Phys.}} \textbf{\bibinfo{volume}{83}}, \bibinfo{pages}{1057} (\bibinfo{year}{2011}).

\bibitem{po2017symmetry}
\bibinfo{author}{Po, H.~C.}, \bibinfo{author}{Vishwanath, A.} \& \bibinfo{author}{Watanabe, H.}
\newblock \bibinfo{title}{Symmetry-based indicators of band topology in the 230 space groups}.
\newblock \emph{\bibinfo{journal}{Nature communications}} \textbf{\bibinfo{volume}{8}}, \bibinfo{pages}{50} (\bibinfo{year}{2017}).

\bibitem{Bradlyn_2017}
\bibinfo{author}{Bradlyn, B.} \emph{et~al.}
\newblock \bibinfo{title}{Topological quantum chemistry}.
\newblock \emph{\bibinfo{journal}{Nature}} \textbf{\bibinfo{volume}{547}}, \bibinfo{pages}{298–305} (\bibinfo{year}{2017}).

\bibitem{Zhang_2019}
\bibinfo{author}{Zhang, T.} \emph{et~al.}
\newblock \bibinfo{title}{Catalogue of topological electronic materials}.
\newblock \emph{\bibinfo{journal}{Nature}} \textbf{\bibinfo{volume}{566}}, \bibinfo{pages}{475–479} (\bibinfo{year}{2019}).

\bibitem{Vergniory_2019}
\bibinfo{author}{Vergniory, M.~G.} \emph{et~al.}
\newblock \bibinfo{title}{A complete catalogue of high-quality topological materials}.
\newblock \emph{\bibinfo{journal}{Nature}} \textbf{\bibinfo{volume}{566}}, \bibinfo{pages}{480–485} (\bibinfo{year}{2019}).

\bibitem{Tang_2019}
\bibinfo{author}{Tang, F.}, \bibinfo{author}{Po, H.~C.}, \bibinfo{author}{Vishwanath, A.} \& \bibinfo{author}{Wan, X.}
\newblock \bibinfo{title}{Comprehensive search for topological materials using symmetry indicators}.
\newblock \emph{\bibinfo{journal}{Nature}} \textbf{\bibinfo{volume}{566}}, \bibinfo{pages}{486–489} (\bibinfo{year}{2019}).

\bibitem{Hohenberg}
\bibinfo{author}{Hohenberg, P.} \& \bibinfo{author}{Kohn, W.}
\newblock \bibinfo{title}{Inhomogeneous electron gas}.
\newblock \emph{\bibinfo{journal}{Phys. Rev.}} \textbf{\bibinfo{volume}{136}}, \bibinfo{pages}{B864--B871} (\bibinfo{year}{1964}).

\bibitem{Kohn}
\bibinfo{author}{Kohn, W.} \& \bibinfo{author}{Sham, L.~J.}
\newblock \bibinfo{title}{Self-consistent equations including exchange and correlation effects}.
\newblock \emph{\bibinfo{journal}{Phys. Rev.}} \textbf{\bibinfo{volume}{140}}, \bibinfo{pages}{A1133--A1138} (\bibinfo{year}{1965}).

\bibitem{Mariette}
\bibinfo{author}{Hellenbrandt, M.}
\newblock \bibinfo{title}{The inorganic crystal structure database (icsd)—present and future}.
\newblock \emph{\bibinfo{journal}{Crystallogr. Rev.}} \textbf{\bibinfo{volume}{10}}, \bibinfo{pages}{17--22} (\bibinfo{year}{2004}).

\bibitem{Watanabe2018mag_SI}
\bibinfo{author}{Watanabe, H.}, \bibinfo{author}{Po, H.~C.} \& \bibinfo{author}{Vishwanath, A.}
\newblock \bibinfo{title}{Structure and topology of band structures in the 1651 magnetic space groups}.
\newblock \emph{\bibinfo{journal}{Science Advances}} \textbf{\bibinfo{volume}{4}}, \bibinfo{pages}{aat8685} (\bibinfo{year}{2018}).

\bibitem{Elcoro2021mag_TQC}
\bibinfo{author}{Elcoro, L.} \emph{et~al.}
\newblock \bibinfo{title}{Magnetic topological quantum chemistry}.
\newblock \emph{\bibinfo{journal}{Nature communications}} \textbf{\bibinfo{volume}{12}}, \bibinfo{pages}{5965} (\bibinfo{year}{2021}).

\bibitem{Peng2022mag_SI}
\bibinfo{author}{Peng, B.}, \bibinfo{author}{Jiang, Y.}, \bibinfo{author}{Fang, Z.}, \bibinfo{author}{Weng, H.} \& \bibinfo{author}{Fang, C.}
\newblock \bibinfo{title}{Topological classification and diagnosis in magnetically ordered electronic materials}.
\newblock \emph{\bibinfo{journal}{Phys. Rev. B}} \textbf{\bibinfo{volume}{105}}, \bibinfo{pages}{235138} (\bibinfo{year}{2022}).
\newblock \urlprefix\url{https://link.aps.org/doi/10.1103/PhysRevB.105.235138}.

\bibitem{samuel1959some}
\bibinfo{author}{Samuel, A.~L.}
\newblock \bibinfo{title}{Some studies in machine learning using the game of checkers}.
\newblock \emph{\bibinfo{journal}{IBM J. Res. Dev.}} \textbf{\bibinfo{volume}{3}}, \bibinfo{pages}{210--229} (\bibinfo{year}{1959}).

\bibitem{zhang2017quantum}
\bibinfo{author}{Zhang, Y.} \& \bibinfo{author}{Kim, E.-A.}
\newblock \bibinfo{title}{Quantum loop topography for machine learning}.
\newblock \emph{\bibinfo{journal}{Phys. Rev. Lett.}} \textbf{\bibinfo{volume}{118}}, \bibinfo{pages}{216401} (\bibinfo{year}{2017}).

\bibitem{zhang2018machine}
\bibinfo{author}{Zhang, P.}, \bibinfo{author}{Shen, H.} \& \bibinfo{author}{Zhai, H.}
\newblock \bibinfo{title}{Machine learning topological invariants with neural networks}.
\newblock \emph{\bibinfo{journal}{Phys. Rev. Lett.}} \textbf{\bibinfo{volume}{120}}, \bibinfo{pages}{066401} (\bibinfo{year}{2018}).

\bibitem{zhang2020interpreting}
\bibinfo{author}{Zhang, Y.}, \bibinfo{author}{Ginsparg, P.} \& \bibinfo{author}{Kim, E.-A.}
\newblock \bibinfo{title}{Interpreting machine learning of topological quantum phase transitions}.
\newblock \emph{\bibinfo{journal}{Phys. Rev. Res.}} \textbf{\bibinfo{volume}{2}}, \bibinfo{pages}{023283} (\bibinfo{year}{2020}).

\bibitem{scheurer2020unsupervised}
\bibinfo{author}{Scheurer, M.~S.} \& \bibinfo{author}{Slager, R.-J.}
\newblock \bibinfo{title}{Unsupervised machine learning and band topology}.
\newblock \emph{\bibinfo{journal}{Phys. Rev. Lett.}} \textbf{\bibinfo{volume}{124}}, \bibinfo{pages}{226401} (\bibinfo{year}{2020}).

\bibitem{2006Compressed}
\bibinfo{author}{Donoho, D.~L.}
\newblock \bibinfo{title}{Compressed sensing}.
\newblock \emph{\bibinfo{journal}{IEEE Transactions on Information Theory}} \textbf{\bibinfo{volume}{52}}, \bibinfo{pages}{1289--1306} (\bibinfo{year}{2006}).

\bibitem{friedman2001greedy}
\bibinfo{author}{Friedman, J.~H.}
\newblock \bibinfo{title}{Greedy function approximation: a gradient boosting machine}.
\newblock \emph{\bibinfo{journal}{Ann. Stat.}} \bibinfo{pages}{1189--1232} (\bibinfo{year}{2001}).

\bibitem{acosta2018analysis}
\bibinfo{author}{Acosta, C.~M.} \emph{et~al.}
\newblock \bibinfo{title}{Analysis of topological transitions in two-dimensional materials by compressed sensing}.
\newblock \emph{\bibinfo{journal}{arXiv}} \bibinfo{pages}{1805.10950} (\bibinfo{year}{2018}).

\bibitem{claussen2020detection}
\bibinfo{author}{Claussen, N.}, \bibinfo{author}{Bernevig, B.~A.} \& \bibinfo{author}{Regnault, N.}
\newblock \bibinfo{title}{Detection of topological materials with machine learning}.
\newblock \emph{\bibinfo{journal}{Phys. Rev. B}} \textbf{\bibinfo{volume}{101}}, \bibinfo{pages}{245117} (\bibinfo{year}{2020}).

\bibitem{cao2020artificial}
\bibinfo{author}{Cao, G.} \emph{et~al.}
\newblock \bibinfo{title}{Artificial intelligence for high-throughput discovery of topological insulators: The example of alloyed tetradymites}.
\newblock \emph{\bibinfo{journal}{Phys. Rev. Mater.}} \textbf{\bibinfo{volume}{4}}, \bibinfo{pages}{034204} (\bibinfo{year}{2020}).

\bibitem{liu2021screening}
\bibinfo{author}{Liu, J.}, \bibinfo{author}{Cao, G.}, \bibinfo{author}{Zhou, Z.} \& \bibinfo{author}{Liu, H.}
\newblock \bibinfo{title}{Screening potential topological insulators in half-heusler compounds via compressed-sensing}.
\newblock \emph{\bibinfo{journal}{J. Phys. Condens. Matter}} \textbf{\bibinfo{volume}{33}}, \bibinfo{pages}{325501} (\bibinfo{year}{2021}).

\bibitem{schleder2021machine}
\bibinfo{author}{Schleder, G.~R.}, \bibinfo{author}{Focassio, B.} \& \bibinfo{author}{Fazzio, A.}
\newblock \bibinfo{title}{Machine learning for materials discovery: Two-dimensional topological insulators}.
\newblock \emph{\bibinfo{journal}{Appl. Phys. Rev.}} \textbf{\bibinfo{volume}{8}}, \bibinfo{pages}{031409} (\bibinfo{year}{2021}).

\bibitem{andrejevic2022machine}
\bibinfo{author}{Andrejevic, N.} \emph{et~al.}
\newblock \bibinfo{title}{Machine-learning spectral indicators of topology}.
\newblock \emph{\bibinfo{journal}{Adv. Mater.}} \textbf{\bibinfo{volume}{34}}, \bibinfo{pages}{2204113} (\bibinfo{year}{2022}).

\bibitem{ma2023topogivity}
\bibinfo{author}{Ma, A.} \emph{et~al.}
\newblock \bibinfo{title}{Topogivity: A machine-learned chemical rule for discovering topological materials}.
\newblock \emph{\bibinfo{journal}{Nano Lett.}} \textbf{\bibinfo{volume}{23}}, \bibinfo{pages}{772--778} (\bibinfo{year}{2023}).

\bibitem{vergniory2022all}
\bibinfo{author}{Vergniory, M.~G.} \emph{et~al.}
\newblock \bibinfo{title}{All topological bands of all nonmagnetic stoichiometric materials}.
\newblock \emph{\bibinfo{journal}{Science}} \textbf{\bibinfo{volume}{376}}, \bibinfo{pages}{eabg9094} (\bibinfo{year}{2022}).

\bibitem{cortes1995support}
\bibinfo{author}{Cortes, C.} \& \bibinfo{author}{Vapnik, V.}
\newblock \bibinfo{title}{Support-vector networks}.
\newblock \emph{\bibinfo{journal}{Mach. Learn.}} \textbf{\bibinfo{volume}{20}}, \bibinfo{pages}{273--297} (\bibinfo{year}{1995}).

\bibitem{Chen_2019}
\bibinfo{author}{Chen, C.}, \bibinfo{author}{Ye, W.}, \bibinfo{author}{Zuo, Y.}, \bibinfo{author}{Zheng, C.} \& \bibinfo{author}{Ong, S.~P.}
\newblock \bibinfo{title}{Graph networks as a universal machine learning framework for molecules and crystals}.
\newblock \emph{\bibinfo{journal}{Chem. Mater.}} \textbf{\bibinfo{volume}{31}}, \bibinfo{pages}{3564–3572} (\bibinfo{year}{2019}).

\bibitem{Dunn_2020}
\bibinfo{author}{Dunn, A.}, \bibinfo{author}{Wang, Q.}, \bibinfo{author}{Ganose, A.}, \bibinfo{author}{Dopp, D.} \& \bibinfo{author}{Jain, A.}
\newblock \bibinfo{title}{Benchmarking materials property prediction methods: the {Matbench} test set and automatminer reference algorithm}.
\newblock \emph{\bibinfo{journal}{npj Comput. Mater.}} \textbf{\bibinfo{volume}{6}} (\bibinfo{year}{2020}).

\bibitem{De_Breuck_2021_1}
\bibinfo{author}{De~Breuck, P.-P.}, \bibinfo{author}{Hautier, G.} \& \bibinfo{author}{Rignanese, G.-M.}
\newblock \bibinfo{title}{Materials property prediction for limited datasets enabled by feature selection and joint learning with modnet}.
\newblock \emph{\bibinfo{journal}{npj Comput. Mater.}} \textbf{\bibinfo{volume}{7}} (\bibinfo{year}{2021}).

\bibitem{xie2018crystal}
\bibinfo{author}{Xie, T.} \& \bibinfo{author}{Grossman, J.~C.}
\newblock \bibinfo{title}{Crystal graph convolutional neural networks for an accurate and interpretable prediction of material properties}.
\newblock \emph{\bibinfo{journal}{Phys. Rev. Lett.}} \textbf{\bibinfo{volume}{120}}, \bibinfo{pages}{145301} (\bibinfo{year}{2018}).

\bibitem{Stone1974}
\bibinfo{author}{Stone, M.}
\newblock \bibinfo{title}{Cross-validatory choice and assessment of statistical predictions}.
\newblock \emph{\bibinfo{journal}{J. R. Stat. Soc., Ser. B, Methodol.}} \textbf{\bibinfo{volume}{36}}, \bibinfo{pages}{111--133} (\bibinfo{year}{1974}).

\bibitem{ho1995random}
\bibinfo{author}{Ho, T.~K.}
\newblock \bibinfo{editor}{{\color{white}a}} (ed.) \emph{\bibinfo{title}{Random decision forests}}.
\newblock (ed.\bibinfo{editor}{{\color{white}a}}) \emph{\bibinfo{booktitle}{Proceedings of 3rd international conference on document analysis and recognition}}, Vol.~\bibinfo{volume}{1}, \bibinfo{pages}{278--282} (\bibinfo{organization}{IEEE}, \bibinfo{year}{1995}).

\bibitem{chen2016xgboost}
\bibinfo{author}{Chen, T.} \& \bibinfo{author}{Guestrin, C.}
\newblock \bibinfo{editor}{{\color{white}a}} (ed.) \emph{\bibinfo{title}{{XGBoost: A Scalable Tree Boosting System} \emph{in} {Proceedings of the 22nd ACM SIGKDD International Conference on Knowledge Discovery and Data Mining}}}.
\newblock (ed.\bibinfo{editor}{{\color{white}a}}) , KDD ’16, \bibinfo{pages}{785–794} (\bibinfo{publisher}{ACM}, \bibinfo{year}{2016}).

\bibitem{De_Breuck_2021_2}
\bibinfo{author}{De~Breuck, P.-P.}, \bibinfo{author}{Evans, M.~L.} \& \bibinfo{author}{Rignanese, G.-M.}
\newblock \bibinfo{title}{Robust model benchmarking and bias-imbalance in data-driven materials science: a case study on modnet}.
\newblock \emph{\bibinfo{journal}{J. Phys. Condens. Matter}} \textbf{\bibinfo{volume}{33}}, \bibinfo{pages}{404002} (\bibinfo{year}{2021}).

\bibitem{van2008visualizing}
\bibinfo{author}{Van~der Maaten, L.} \& \bibinfo{author}{Hinton, G.}
\newblock \bibinfo{title}{Visualizing data using {t-SNE}}.
\newblock \emph{\bibinfo{journal}{J. Mach. Learn. Res.}} \textbf{\bibinfo{volume}{9}} (\bibinfo{year}{2008}).

\bibitem{Jain}
\bibinfo{author}{Jain, A.} \emph{et~al.}
\newblock \bibinfo{title}{{Commentary: The Materials Project: A materials genome approach to accelerating materials innovation}}.
\newblock \emph{\bibinfo{journal}{APL Mater.}} \textbf{\bibinfo{volume}{1}}, \bibinfo{pages}{011002} (\bibinfo{year}{2013}).

\bibitem{ONG2013314}
\bibinfo{author}{Ong, S.~P.} \emph{et~al.}
\newblock \bibinfo{title}{Python materials genomics (pymatgen): A robust, open-source python library for materials analysis}.
\newblock \emph{\bibinfo{journal}{Comput. Mater. Sci.}} \textbf{\bibinfo{volume}{68}}, \bibinfo{pages}{314--319} (\bibinfo{year}{2013}).

\bibitem{Ward2017}
\bibinfo{author}{Ward, L.} \emph{et~al.}
\newblock \bibinfo{title}{{Including crystal structure attributes in machine learning models of formation energies via Voronoi tessellations}}.
\newblock \emph{\bibinfo{journal}{Phys. Rev. B}} \textbf{\bibinfo{volume}{96}}, \bibinfo{pages}{024104} (\bibinfo{year}{2017}).
\newblock \urlprefix\url{http://link.aps.org/doi/10.1103/PhysRevB.96.024104}.

\bibitem{Ward2016}
\bibinfo{author}{Ward, L.}, \bibinfo{author}{Agrawal, A.}, \bibinfo{author}{Choudhary, A.} \& \bibinfo{author}{Wolverton, C.}
\newblock \bibinfo{title}{A general-purpose machine learning framework for predicting properties of inorganic materials}.
\newblock \emph{\bibinfo{journal}{npj Comput. Mater.}} \textbf{\bibinfo{volume}{2}} (\bibinfo{year}{2016}).

\bibitem{Deml}
\bibinfo{author}{Deml, A.~M.}, \bibinfo{author}{O'Hayre, R.}, \bibinfo{author}{Wolverton, C.} \& \bibinfo{author}{Stevanovi\ifmmode~\acute{c}\else \'{c}\fi{}, V.}
\newblock \bibinfo{title}{Predicting density functional theory total energies and enthalpies of formation of metal-nonmetal compounds by linear regression}.
\newblock \emph{\bibinfo{journal}{Phys. Rev. B}} \textbf{\bibinfo{volume}{93}}, \bibinfo{pages}{085142} (\bibinfo{year}{2016}).
\newblock \urlprefix\url{https://link.aps.org/doi/10.1103/PhysRevB.93.085142}.

\bibitem{WARD201860}
\bibinfo{author}{Ward, L.} \emph{et~al.}
\newblock \bibinfo{title}{Matminer: An open source toolkit for materials data mining}.
\newblock \emph{\bibinfo{journal}{Comput. Mater. Sci.}} \textbf{\bibinfo{volume}{152}}, \bibinfo{pages}{60--69} (\bibinfo{year}{2018}).

\bibitem{Olson2016EvoBio}
\bibinfo{author}{Olson, R.~S.} \emph{et~al.}
\newblock \emph{\bibinfo{title}{Applications of Evolutionary Computation: 19th European Conference, EvoApplications 2016, Porto, Portugal, March 30 -- April 1, 2016, Proceedings, Part I}}, Ch. \bibinfo{chapter}{Automating Biomedical Data Science Through Tree-Based Pipeline Optimization}, \bibinfo{pages}{123--137} (\bibinfo{publisher}{Springer International Publishing}, \bibinfo{year}{2016}).

\bibitem{Faber2015}
\bibinfo{author}{Faber, F.}, \bibinfo{author}{Lindmaa, A.}, \bibinfo{author}{von Lilienfeld, O.~A.} \& \bibinfo{author}{Armiento, R.}
\newblock \bibinfo{title}{Crystal structure representations for machine learning models of formation energies}.
\newblock \emph{\bibinfo{journal}{Int. J. Quantum Chem.}} \textbf{\bibinfo{volume}{115}}, \bibinfo{pages}{1094--1101} (\bibinfo{year}{2015}).

\bibitem{miedema_zhang_2016}
\bibinfo{author}{Zhang, R.}, \bibinfo{author}{Zhang, S.}, \bibinfo{author}{He, Z.}, \bibinfo{author}{Jing, J.} \& \bibinfo{author}{Sheng, S.}
\newblock \bibinfo{title}{Miedema calculator: A thermodynamic platform for predicting formation enthalpies of alloys within framework of miedema’s theory}.
\newblock \emph{\bibinfo{journal}{Comput. Phys. Commun.}} \textbf{\bibinfo{volume}{209}}, \bibinfo{pages}{58--69} (\bibinfo{year}{2016}).

\bibitem{Svetlana}
\bibinfo{author}{Kotochigova, S.}, \bibinfo{author}{Levine, Z.~H.}, \bibinfo{author}{Shirley, E.~L.}, \bibinfo{author}{Stiles, M.~D.} \& \bibinfo{author}{Clark, C.~W.}
\newblock \bibinfo{title}{Local-density-functional calculations of the energy of atoms}.
\newblock \emph{\bibinfo{journal}{Phys. Rev. A}} \textbf{\bibinfo{volume}{55}}, \bibinfo{pages}{191--199} (\bibinfo{year}{1997}).

\bibitem{Talirz2020}
\bibinfo{author}{Talirz, L.} \emph{et~al.}
\newblock \bibinfo{title}{Materials cloud, a platform for open computational science}.
\newblock \emph{\bibinfo{journal}{Scientific Data}} \textbf{\bibinfo{volume}{7}} (\bibinfo{year}{2020}).
\newblock \urlprefix\url{http://dx.doi.org/10.1038/s41597-020-00637-5}.

\bibitem{Fraux2020}
\bibinfo{author}{Fraux, G.}, \bibinfo{author}{Cersonsky, R.} \& \bibinfo{author}{Ceriotti, M.}
\newblock \bibinfo{title}{Chemiscope: interactive structure-property explorer for materials and molecules}.
\newblock \emph{\bibinfo{journal}{Journal of Open Source Software}} \textbf{\bibinfo{volume}{5}}, \bibinfo{pages}{2117} (\bibinfo{year}{2020}).
\newblock \urlprefix\url{http://dx.doi.org/10.21105/joss.02117}.

\end{thebibliography}

\begin{appendices}
\section*{Appendix}
\setcounter{figure}{0}
\setcounter{table}{0}
\renewcommand{\thefigure}{A\arabic{figure}}
\renewcommand{\thetable}{A\arabic{table}}

\begin{table}[]
	\centering
	\caption{$F_1$ score, precision, and recall (in \%) of the different nested cross-validation (NCV) and generalization tests depending on the training dataset.
    The results on (part of) the training dataset evaluated by NCV are indicated by a star.}
    \setlength{\tabcolsep}{6pt}
    \begin{tabular}{llccccccc} \hline \hline
    	\multicolumn{2}{l}{       Test   \textbackslash{}   Train}  
        & $M$ & $T$ & $M\backslash T$ & $M\cap T$  & $T\backslash M$ & $(M\backslash T)\cup(T\backslash M)$  & $M\cup T$ \\
        \hline
    	\multirow{5}{*}{$F_1$ score}
        & $M\backslash T$ & 94.9* & 93.4* & 94.4\0& 94.0* & 90.6\0& 95.2\0& 94.9* \\ 
    	& $M\cap T$       & 92.7* & 92.7\0& 87.6* & 92.9\0& 88.0\0& 88.9* & 92.6* \\  
    	& $T\backslash M$ & 91.1\0& 90.1* & 88.7\0& 91.2\0& 89.1* & 90.1* & 90.9* \\ 
    	& NCV             & 93.7\0& 91.6\0& 94.4\0& 92.7\0& 89.4\0& 92.6\0& 92.9\0 \\ 
    	& $M\cup T$       & 93.0\0& 92.2\0& 90.3\0& 92.8\0& 89.2\0& 91.4\0& 92.9 \0\\ 
        \hline
    	\multirow{5}{*}{Precision}
        & $M\backslash T$ & 83.7* & 81.7* & 84.0\0& 77.4* & 82.5\0& 86.1\0& 85.3* \\ 
    	& $M\cap T$       & 85.3* & 84.8\0& 84.9* & 82.1\0& 83.5\0& 86.4* & 86.7* \\
    	& $T\backslash M$ & 63.7\0& 70.2* & 64.3\0& 58.7\0& 69.7* & 71.3* & 72.5* \\
    	& NCV             & 84.4\0& 77.4\0& 84.0\0& 81.9\0& 69.6\0& 78.8\0& 81.2\0\\ 
    	& $M\cup T$       & 77.5\0& 78.8\0& 77.7\0& 72.7\0& 78.5\0& 81.3\0& 81.5\0\\
        \hline
    	\multirow{5}{*}{Recall} 
        & $M\backslash T$ & 88.9* & 87.2* & 88.9\0& 84.9* & 86.4\0& 90.4\0& 89.8* \\ 
    	& $M\cap T$       & 88.8* & 88.6\0& 86.2* & 87.2\0& 85.7\0& 87.6* & 89.5* \\ 
    	& $T\backslash M$ & 75.0\0& 78.9* & 74.5\0& 71.4\0& 78.2* & 79.6* & 80.7* \\
    	& NCV             & 88.8\0& 83.9\0& 88.9\0& 87.0\0& 78.3\0& 85.1\0& 86.7\0\\ 
    	&$M\cup T$        & 84.6\0& 85.0\0& 83.5\0& 81.5\0& 83.5\0& 86.0\0& 86.8\0\\
        \hline \hline
    \end{tabular}
    \label{tab:train_test_results2}
\end{table}
 
\begin{figure}
  	\centering
   	\includegraphics[width=1\textwidth]{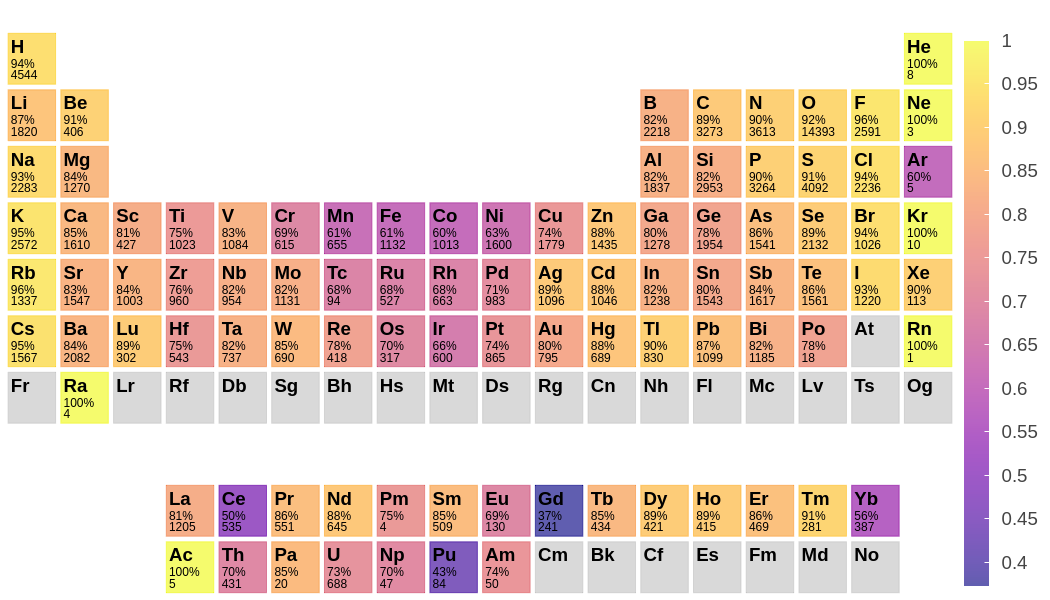}
   	\caption{Periodic table with the average accuracy of the XGBoost model in the dataset $M\cup T$ per element.
    For each element, two numbers are reported.
    The first one (in \%) shows the average accuracy for the compounds containing the element.
    This number is also reflected in the coloring of the cell in the periodic table.
    The second on illustrates the number of compounds containing this element.}
   	\label{fig:elems_acc}
\end{figure}

\begin{table}[]
	\centering
	\caption{Comparison of the accuracy, F1 score, precision, and recall (in \%) on the datasets $M\backslash T$, $M\cap T$ and $T\backslash M$ with and without the materials containing at least one of the following elements: Ne, Mn, Fe, Eu, Gd, Po, Rn, Ra, Am.
    The first three columns show the NCV results on each sub-dataset, while the last three columns are the NCV results on the dataset $M\cup T$.
    The proportion of the problematic elements is also reported in the last row of the table for each dataset.
    }
      \setlength{\tabcolsep}{8pt}
	\begin{tabular}{llccccccc}
    \hline
    \hline
	\multicolumn{2}{c}{\multirow{2}{*}{}} & \multicolumn{3}{c}{NCV on each sub-dataset} && \multicolumn{3}{c}{NCV on $M\cup T$} \\
    \cmidrule{3-5} \cmidrule{7-9}
	\multicolumn{2}{c}{}       & $M\backslash T$ & $M\cap T$ & $T\backslash M$ &&  $M\backslash T$ & $M\cap T$ & $T\backslash M$ \\
    \hline
	\multirow{2}{*}{Accuracy}
    & with    & 84.1 & 85.4 & 72.0 && 85.9 & 86.6 & 74.2 \\
	& without & 84.7 & 85.9 & 75.1 && 86.3 & 87.2 & 77.5 \\
    \hline
	\multirow{2}{*}{$F_1$ score}
    & with    & 88.9 & 87.2 & 78.2 && 89.8 & 89.5 & 80.7 \\
	& without & 89.0 & 87.2 & 78.8 && 89.7 & 89.6 & 81.8 \\
    \hline
	\multirow{2}{*}{Precision}
    & with    & 94.4 & 92.9 & 89.1 && 94.9 & 92.6 & 90.9 \\
	& without & 94.7 & 93.1 & 89.8 && 95.1 & 92.8 & 91.6 \\
    \hline
	\multirow{2}{*}{Recall}
    & with    & 84.0 & 82.1 & 69.7 && 85.3 & 86.7 & 72.5 \\
	& without & 84.0 & 82.0 & 70.1 && 85.0 & 86.7 & 73.9 \\
    \hline
	\multicolumn{2}{l}{Element proportion (in \%)}
    & 3.1  & 1.9  & 15.9 && 3.1    & 1.9    & 15.9  \\
    \hline
    \hline
   \end{tabular}
   \label{tab:score_elel_proportion_in1}
\end{table}

\begin{table}[]
    \centering
	\caption{Comparison of the NCV accuracy, $F_1$ score, precision and recall (in \%) on the dataset $M\cup T$ with and without materials containing Cr, Mn, Fe, Cu, Tc, Eu, Os, and Np.}
    \setlength{\tabcolsep}{9pt}
	\begin{tabular}{llcccccccc}
    \hline
    \hline
		\multicolumn{2}{l}{}    & Cr     & Mn    & Fe    & Cu  & Tc  & Eu   & Os & Np  \\
        \hline
		\multirow{2}{*}{Accuracy}   & with   & 69.3 & 61.4 & 60.6 & 74.3 & 68.1 & 69.2 & 70.0 & 70.2  \\
		& without   &  83.2 & 83.3 & 83.7 & 83.4 & 83.0 & 83.0 & 83.1 & 83.0    \\
        \hline
		\multirow{2}{*}{$F_1$ score}    & with   & 85.1 & 78.5 & 79.6 & 81.0 & 86.0 & 72.4 & 83.3 & 89.3  \\

		& without   & 86.7 & 87.0 & 87.1 & 87.1 & 86.7 & 86.8 & 86.7 & 86.7  \\
        \hline
		\multirow{2}{*}{Precision}   & with   & 92.6 & 90.6 & 88.0 & 94.4 & 93.0 & 90.5 & 89.1 & 100.0 \\

		& without   & 93.0 & 93.0 & 93.2 & 92.8 & 92.9 & 93.0 & 93.0 & 92.9  \\
        \hline
		\multirow{2}{*}{Recall}   & with   & 78.7 & 69.3 & 72.6 & 70.9 & 80.0 & 60.3 & 78.2 & 80.6  \\

		& without   & 81.3 & 81.7 & 81.7 & 82.1 & 81.2 & 81.4 & 81.3 & 81.2 \\
     \hline
     \hline
	\end{tabular}
	\label{tab:score_elel_acc}
\end{table}

\begin{table}[]
	\centering
	\caption{Comparison of the NCV accuracy, $F_1$ score, precision and recall (in \%) on the whole dataset $T\backslash M$ with the same results on the subsets 1 obtained by excluding the materials containing selected elements (Cr, Mn, Fe, Cu, Tc, Eu, Os or Np), and subset 2 constructed by further excluding materials without MP-ID.}
    \setlength{\tabcolsep}{17pt}
	\begin{tabular}{lccccc}
    \hline
    \hline
		\textbf{}     & Count & Accuracy    & $F_1$ score     & Precision    & Recall      \\
        \hline
		All data  & 9,925 & 72.0 & 78.2 & 89.1 & 69.7 \\
		Subset 1  & 7,364 & 77.3 & 79.9 & 90.1 & 71.8 \\
        Subset 2  & 3,550 & 78.3 & 82.9 & 90.4 & 76.6 \\ 
    \hline
    \hline
	\end{tabular}
	\label{tab:score_elel_mpid1}
\end{table}
    
\begin{table}[]
	\centering
	\caption{NCV accuracy, $F_1$ score, precision and recall (in \%) on the datasets $\widetilde{M\backslash T}$, $\widetilde{M\cap T}$ and $\widetilde{T\backslash M}$.} 
    \setlength{\tabcolsep}{22.5pt}
	\begin{tabular}{lcccc}
    \hline
    \hline
		\textbf{}  & Accuracy    & $F_1$ score    & Precision    & Recall    \\ \hline\\[-1.7ex]
		$\widetilde{M\backslash T}$     & 79.2 & 82.8 & 92.8 & 74.7 \\[3pt]
		$\widetilde{M\cap T}$           & 79.9 & 84.4 & 90.7 & 78.8 \\[3pt]
		$\widetilde{T\backslash M}$     & 77.3 & 78.4 & 88.1 & 70.7 \\
    \hline
    \hline
	\end{tabular}
	\label{tab:ncv_sample}
\end{table}
    
\begin{table}[]
\centering
\caption{Comparison of the value of $g(M)$ (see Eq.~{eq:topog}) based on the elemental topogivity $\tau_E$ from this work and from Ref.~\cite{ma2023topogivity} for well-known nontrivial topological materials.
Their type as determined experimentally is also reported.
The ``*'' indicated for the type of TaAs refers to the fact that it is a Weyl semimetal which cannot be identified by symmetry-indicator theory nor TQC.
CoSi is an example where Ref.~\cite{ma2023topogivity} does not allow for a prediction because the topogivity of Co is not available.} 
\setlength{\tabcolsep}{29pt}
\begin{tabular}{lccc}
\hline
\hline
Compound & Type & This work & Ref.~\protect\cite{ma2023topogivity} \\
\hline
Cd$_2$As$_3$ & HSLSM & -2.815 & -5.375 \\
CoSi         & HSPSM &\phantom{-}3.586 &  ---   \\
Na$_3$Bi     & HSLSM & -1.184 & \phantom{-}1.606 \\
PtAl         & HSPSM &\phantom{-}2.466 & \phantom{-}3.023 \\
SnTe         & TCI   & -1.219 & -0.574 \\
TaAs$_2$     & TCI   &\phantom{-}0.248 & \phantom{-}0.410 \\
Bi$_2$Se$_3$ & TI    & -5.185 & -4.411 \\
Bi$_2$Te$_3$ & TI    & -2.959 & -1.102 \\
Sb$_2$Te$_3$ & TI    & -5.965 & -4.118 \\
ZrTe$_5$     & TI    & -7.440 & -5.816 \\
TaAs         & *     &\phantom{-}1.711 & \phantom{-}2.899 \\    
\hline
\hline
\end{tabular}
\label{tab:topogivity_test}
\end{table}

\end{appendices}

\end{document}